\documentclass[fleqn,usenatbib]{mnras}
\usepackage{booktabs}       % in preamble
\usepackage{threeparttable} % in preamble
\usepackage{newtxtext,newtxmath}
\usepackage[T1]{fontenc}
\usepackage[normalem]{ulem} % for \sout, \add, \del

\newcommand{\taya}[1]{\textcolor{black}{#1}}
\newcommand{\fo}{f_{\rm ox}}
\newcommand{\yi}{y_{\rm in}}
\newcommand{\tTin}{\tau_\text{T,in}}
\newcommand{\tT}{\tau_\text{T}}
\newcommand{\tabs}{\tau_\text{abs}}
\newcommand{\xiten}{\xi_\text{10keV}}
\newcommand{\xiCten}{\xi_\text{Comp,10}}
\newcommand{\xiC}{\xi_\text{Comp}}
\newcommand{\Sigfit}{\Sigma_\text{fit,nt}}
\newcommand{\TC}{T_\text{Comp}}
\newcommand{\kB}{k_\text{B}}

\DeclareRobustCommand{\VAN}[3]{#2}
\let\VANthebibliography\thebibliography
\def\thebibliography{\DeclareRobustCommand{\VAN}[3]{##3}\VANthebibliography}
\usepackage{array}
\newcolumntype{P}[1]{>{\centering\arraybackslash}p{#1}}

\usepackage{graphicx}	% Including figure files
\usepackage{amsmath}	% Advanced maths commands

\title[X-ray Transmission]{X-ray Transmission Through Photoionized Gas with Moderate Thomson Optical Depth}

\author[Govreen-Segal et al.]{
Taya Govreen-Segal,$^{1}$\thanks{E-mail: taya@govreensegal.com}
Ehud Nakar,$^{1}$
and Eliot Quataert$^{2}$
\\
$^{1}$School of Physics and Astronomy, Tel Aviv University, Tel Aviv 6997801, Israel\\
$^{2}$Department of Astrophysical Sciences, Princeton University,
Princeton, NJ 08544, USA\\
}

\pubyear{\the\year{}}

\begin{document}
\label{firstpage}
\pagerange{\pageref{firstpage}--\pageref{lastpage}}
\maketitle

\begin{abstract}
We model the absorption of X-rays by gas obscuring the source and photoionized by it. We consider a broad range of column densities, including both Thomson-thin and Thomson-thick media. For the Thomson thin regime, we derive a simple criterion based on the source luminosity and spectrum, as well as the medium radius and column density, that distinguishes between the following cases: (i) The absorption can be modeled well by a neutral medium; (ii) The radiation ionizes its way through the medium, and no absorption is expected; and (iii) A detailed model is required because the column density inferred from modeling the absorption with a neutral gas is much lower than the actual column density, or because the absorption features cannot be fitted by a neutral absorber. We derive the criterion analytically using a toy model of hydrogen and oxygen and calibrate it for realistic compositions with metallicities in the range $Z/Z_{\odot}=0.01-50$, using \textsc{Cloudy}.  
We generalize the model to the Thomson-thick regime, where we consider, alongside photoabsorption, electron scattering, Compton heating, Comptonization, and photon degradation. In this case, the emergent spectrum depends on the boundary condition experienced by photons scattered back towards the source. We discuss the effect of a reflective boundary and a reprocessing boundary. We provide simple criteria for the expected absorption state and discuss additional effects that alter the spectrum. The main motivation for our modeling is X-ray emission from supernovae interacting with the circumstellar medium; however, we expect it to be useful for many other applications.
\end{abstract}

% % Select between one and six entries from the list of approved keywords.
% % Don't make up new ones.

%%%%%%%%%%%%%%%%%%%%%%%%%%%%%%%%%%%%%%%%%%%%%%%%%%

%%%%%%%%%%%%%%%%% BODY OF PAPER %%%%%%%%%%%%%%%%%%

\section{Introduction}
The observed X-ray spectrum from a source obscured by an absorbing medium depends both on the intrinsic source emission and on attenuation by the intervening gas. If the medium has a sufficiently high column density, the absorption can strongly modify the observed spectrum. Or, if the source is bright enough to ionize the intervening matter, the medium may become transparent and produce little to no absorption. As the absorption depends sensitively on the ionization state of the gas, this state and its effect on the spectrum must be modeled in order to recover the source SED. Such modeling also enables measurements of the intervening column density and predictions of the reprocessed emission. This interplay between high energy emission by an X-ray source and the ionization of, and absorption by, intervening matter is important in a wide range of astrophysical systems, including supernovae \citep{Nymark2006}, fast blue optical transients \citep[e.g.,][]{Margutti2019,Nayana2025b}, \taya{tidal disruption events \citep[e.g.,][]{Roth2015,Auchettl2017,Dai2018}, GRB afterglows}, and active galactic nuclei \citep{Netzer2008}.

A model that is often used in X-ray spectral analysis is the neutral absorber model, e.g., the one by \cite{Wilms2000}, which is implemented in \textsc{xspec} \citep{XSPEC} modules such as \texttt{tbabs}. This model assumes the intervening matter is entirely neutral, \taya{and is used to report the "neutral equivalent" column density, which is a lower limit on the actual column density}. \taya{When the metals are not highly ionized, and the metallicity of the absorbing matter is known, this measurement is a good approximation for the actual column density. However,} a sufficiently bright source \taya{illuminating a column density dominated by matter in the vicinity of the source, can lead to a significant fraction of the column density being ionized.} \taya{This either produces a neutral equivalent column density that significantly underestimates the actual column density, or } produces absorption features that cannot be fit with a neutral absorber. \taya{While various ionized absorber or warm absorber models exist for fitting X-ray spectra, these typically require prior knowledge of the temperature of the medium, information that is typically unavailable.}

Currently, in many setups, the only way to safely estimate the absorbed and transferred radiation is through a full photoionization simulation using a code such as \textsc{Cloudy} \citep[][]{CLOUDY2023} or \textsc{xstar} \citep{XSTAR}. These simulations calculate the non-local thermal equilibrium (NLTE) gas ionization state and can provide accurate predictions of both absorption and reprocessed emission. Yet, given their complexity, they are not always practical for routine analysis of observations or for use as part of broader analytical works. It is therefore useful to have a simple criterion that distinguishes between three regimes: (i) when the neutral absorber assumption remains adequate (possibly with calibration), (ii) when the gas is fully ionized and absorption is negligible, and (iii) when detailed simulations are required. Establishing such a criterion is the first goal of this paper.

A major limitation of these codes is that they cannot be used if the intervening medium is Thomson-thick (i.e., optically thick to electron scattering). Thus, a second goal of this paper is to \taya{extend this criterion to} address the Thomson-thick regime. In this regime, standard photoionization simulations are not valid because they do not correctly treat electron scattering. More sophisticated Monte-Carlo radiative transfer simulations are required, but these are computationally expensive and not widely accessible. Several such simulations exist \citep[see][and references therein]{Ross2005}, but most are not publicly available; to the best of our knowledge, the exceptions are \textsc{MOCASSIN} \citep{Ercolano2003}, which is not applicable to all the cases considered here, and \textsc{Sedona} \citep{Kasen2006,Roth2015}, which includes most of the relevant physics but does not yet include high ionization degrees, and Compton \taya{effects}. This motivates the need for analytical methods to approximate the emergent spectrum in Thomson-thick environments.

In Thomson-thin media, the matter is photoionized by the incoming radiation, and any ion that is not fully ionized contributes to the absorption of the incident spectrum. In Thomson-thick media, however, additional complications arise. X-ray photons spend more time in the medium as they scatter multiple times through the same part of the gas, which produces two competing effects: increased absorption probability for each photon and increased ionization of the gas due to the higher ionizing photon density. Both effects, along with the proper treatment of photons scattered back toward the source, must be considered. Furthermore, Compton degradation, Comptonization of lower-energy photons into the X-ray observing range, and Compton heating play a significant role in shaping the emergent spectrum.

Several semi-analytical and numerical studies have investigated absorption in both Thomson-thin and Thomson-thick cases. However, many rely on simplifying assumptions that are not always valid or are based on a limited set of simulations used to explore ionization profiles \citep[][]{Tarter1969A,Tarter1969B,Hatchett1976,Krolik1984,Metzger2014}. More detailed modeling exists for accretion disks, but it primarily focuses on reflected X-rays, i.e., X-rays scattered back towards the source \citep[see][and references therein]{Ross2005}. Importantly, none of these studies provide a clear criterion for when the neutral absorber assumption is valid or when the gas is fully ionized and absorption can be neglected.

While X-ray absorption is relevant in many astrophysical environments, our study is primarily motivated by supernovae (SNe) interacting with dense circumstellar material (CSM). In these systems, measuring the column density of the CSM is of special interest since it probes the mass loss rate of the pre-explosion star. Due to the complexity of the system, theoretical studies addressing the Thomson-thin CSM and including treatment of ionization are typically limited to case-by-case numerical investigations \citep[e.g.,][]{Chevalier1994,Nymark2006}. Meanwhile, the lack of tools for treating photoionization in Thomson-thick media has led to models that either neglect photoabsorption \taya{(an assumption that, in light of the findings of this paper, is not always correct), or use approximations that are also inconsistent with our findings here} \citep{Svirski2012,Svirski2014, Margalit2022,Wasserman2025}. \taya{Other works rely on a small number of \textsc{Cloudy} simulations, and the published results of older simulations that are not self-consistent, to point out when photoabsorption may be important \citep{Chevalier2012}, but do not treat it in their model.} 

% \taya{\cite{Margalit2022} use a stromgren sphere approximation, which assumes that recombination depletes the reservoir of reis invalid for treating high ionization states of ions, since  \cite{Wasserman2025}, who argue that "the energy required for ionization is dominated by the hydrogen atoms", and conclude that since the hydrogen is fully ionized, the metals are fully ionized too}.
The criteria we aim to establish are particularly important for interacting SNe, as the ionization state of the CSM can vary significantly across the relevant parameter space. Even within a single event, as time progresses, the gas may transition between neutral and fully ionized states or vise versa. Consequently, our numerical tests focus on the phase space relevant to SNe–CSM interaction, and we normalize results to canonical values representative of these systems.

In the following sections, we start by treating the Thomson-thin regime (\S\ref{sec: Thomson Thin}), where we derive an analytical model and validate it numerically using \textsc{Cloudy} simulations\taya{, and discuss the application of this criterion to line emission and absorption}. We then extend the analysis to the Thomson-thick regime (\S\ref{sec: Thomson thick}), incorporating various boundary conditions and additional physical processes that influence the emergent spectrum. In section \S\ref{sec:examples}, we provide two examples of how to use our model by applying it to the observations of SN\,2023ixf and SN\,2008D. Finally, we summarize our findings in \S\ref{sec:summary}.

Table \ref{tab:W} summarizes the criteria for the expected absorption for the Thomson thin medium, and Tables \ref{tab:reflective} and \ref{tab:reprossesing} do so for the Thomson thick case, for reflective and reprocessing boundaries, respectively. These tables are intended to be used without detailed knowledge of the rest of the paper.

%%%%%%%%%%%%%%%%%%%%%%%%%%%%%%%%%%%%%%%%%%%%%%%%%%

\section{X-ray absorption in Thomson thin media}\label{sec: Thomson Thin}
Consider an X-ray source that illuminates a slab of matter. The transmitted X-ray spectrum depends on the frequency-dependent cross section of the absorbing matter, which is determined by its composition and ionization state. The ionization state is set by the ionization and recombination rates, which, in turn, depend, among other factors, on the X-ray transmission through the medium. Therefore, the X-ray radiative transfer and the gas ionization state must be modeled self-consistently.

Considering a Thomson-thin medium, i.e., matter with electron column density $\Sigma \lesssim 1.5 \cdot 10^{24}~\text{cm}^{-2}$, simplifies the problem for a couple of reasons. First, the approximation that the only important processes are photoionization and radiative recombination is valid. Namely, we can neglect electron scattering, Comptonization, and Compton degradation, which are unimportant in the Thomson thin regime. We also neglect Compton heating and collisional ionization in our analytical model since we later find, using simulations, that they are not important for the criterion we derive. We further discuss this in Appendix \ref{App:collisional_ion}. Second, in the Thomson-thin regime, we can use \textsc{Cloudy} to validate and calibrate the analytical model.  

We limit our discussion to X-ray sources in which the number of X-ray photons is dominated by low energies, e.g., specific luminosities $L_{\nu} \propto \nu^{\alpha}$ where $\alpha \leq 0$, such as a free-free spectrum, and in which induced Compton heating is either negligible relative to ordinary Compton heating\footnote{The criterion that induced Compton heating is negligible compared to ordinary Compton heating is $h\int \nu L_{\nu} \, d\nu \gg \frac{c^{2}}{32\pi}\frac{\int \nu^{-2} L_{\nu}^{2} \, d\nu}{4\pi r^{2} }$}, or else does not lead to a Compton temperature above $10^7$ K. In the numerical section, we test spectra with $-2 \leq \alpha \leq 0$, and while we expect the results to be valid also at smaller values of $\alpha$, this has not been verified. We also limit our model to column densities $\Sigma \gtrsim 1.2 \cdot 10^{21}~\text{cm}^{-2} \left(\frac{Z}{Z_{\odot}}\right)^{-1}$, so that the column density of the metals is sufficient to cause significant absorption at $\sim 1 {\rm ~keV}$ when the metals are neutral. We consider only compositions in which hydrogen is abundant and assume that hydrogen is fully ionized, either directly by radiation or by secondary electrons. All our simulations fall into this regime, and we do not model the transition to the regime where the entire medium is neutral.

Since we do not consider relativistic effects, we also limit the cutoff temperature of the spectrum to a few times $10^{9}$~K. We also limit our discussion to densities at which the recombination coefficient is roughly independent of electron density\footnote{\label{fn:three-body}While it is accepted that above $\sim 10^{12}~\text{cm}^{-3}$, three-body recombination becomes important and $\alpha_{\text{B}}$ becomes density dependent; that rule of thumb is based on the assumption of matter at $\sim 10^{4}$~K. We find that, in our case, the higher temperatures push this transition to higher densities. This can also be seen in the values presented in \cite{Storey1995}.}, $n \lesssim 10^{15}~\text{cm}^{-3}$. We also restrict ourselves to media with density profiles in which most of the column density arises from matter near the inner radius, $r_{\text{in}}$, such as uniform-density matter with a width of $\Delta r\lesssim r$, or a wind profile (i.e., $n \propto r^{-2}$), and assume a steady-state solution, i.e., that the source properties don't change significantly on time scales shorter than the time it takes to establish ionization equilibrium:
\begin{equation}
    t_{eq}\approx 800 {\rm~sec} \left(\frac{n}{10^8 {\rm cm^{-3}}}\right)^{-1}\left(\frac{T}{10^5{\rm ~K}}\right),
\end{equation}
\taya{where $n$ and $T$ are the density and temperature of the region of interest. 
This timescale is derived and discussed in Appendix \ref{App:steadystate})}. Finally, we do not \taya{precisely model} line emission and absorption features, such as iron K$\alpha$, which may affect a specific frequency range in the spectrum\taya{, and only briefly discuss the consequences of our model for emission and absorption features, in \S\ref{sec:lines}}. We discuss various extensions to our model, such as the effect of relaxing the assumptions about the spectrum and the effect of induced Compton heating in Appendix~\ref{App:extensions}.

\subsection{Analytical model}
Let us first consider a toy model of only hydrogen and oxygen, where the number fraction of oxygen is $\fo \ll 1$, and hydrogen is fully ionized\footnote{While for a hard spectrum, photoionization of hydrogen and helium may be inefficient, the ionization of metals heats the matter, causing lighter elements to be collisionally ionized, and also releases super-thermal electrons that efficiently collisionally ionize the hydrogen. See the discussion in \cite{Draine2011} on secondary ionization efficiency.}, so that the electron number density is independent of the oxygen ionization state. This model enables us to analytically derive basic results before generalizing to more realistic abundances.  Since oxygen is typically the main absorber at energies $\sim 1~{\rm keV}$ \citep[see figure 1 of][]{Wilms2000}, this model should be a good approximation for such energies, also for abundance ratios similar to solar. This expectation is verified using \textsc{Cloudy} simulations in the following section.

Consider the following setup. An X-ray source illuminates a slab of matter with an inner radius  $r_\text{in}$ and a column density $\Sigma$. Since the absorption cross-section of oxygen at $\sim 1~{\rm keV}$ energies depends primarily on the presence of K-shell electrons and is barely altered by additional bound electrons in outer shells, the absorption cross-section of the matter depends only on the fraction of the oxygen that is not fully ionized (which will be derived in the following paragraphs). Thus, the total optical depth for absorption at a given frequency in the keV range is 
\begin{equation}
    \tabs(h\nu\sim {\rm 1~ keV}) \approx \int y \fo n_e \sigma_{O7\to 8}(\nu) {\rm dr}
\end{equation}
where $n_e$ is the electron number density, $\sigma_{O7\to 8}$ is the absorption cross section of oxygen in the $O^{7+}$ state, which is similar to the cross section of neutral oxygen above $0.87~{\rm keV}$. $y$ is the fraction of oxygen ions that are not fully ionized, given by:
\begin{equation}
    y\equiv\frac{n_{O}-n_{O^{8+}}}{n_{O}}
\end{equation}
where $n_{O^{X+}}$ is the number density of $X$ times ionized oxygen, i.e., $n_{O^{8+}}$ is the number density of fully ionized oxygen, and $n_O$ is the total oxygen density.

\begin{figure}
    \centering
    \includegraphics[width=\columnwidth]{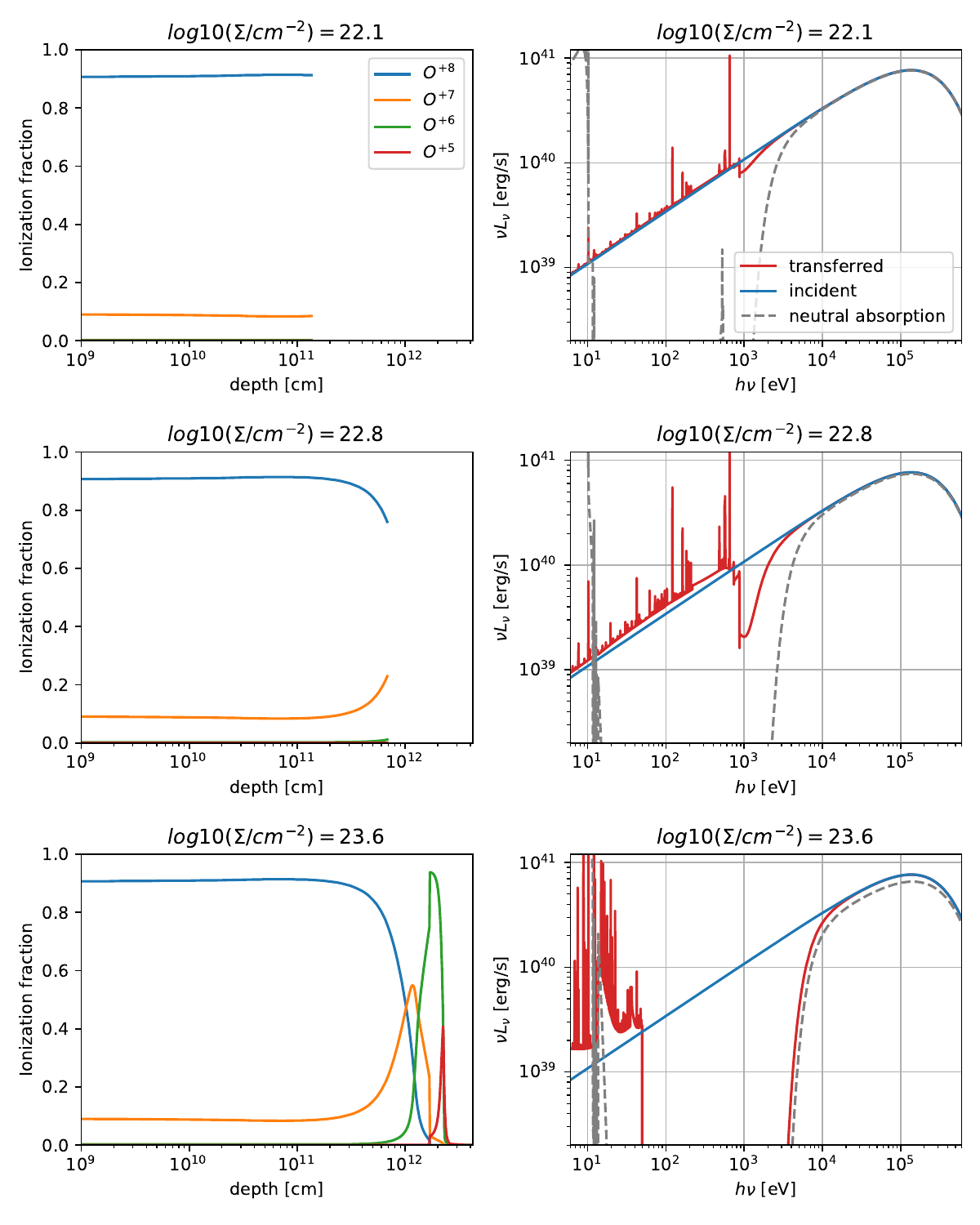}
    \caption{The ionization profile {\it (left panels)} and transmitted spectrum {\it (right panels)} of simulations with   $1-\yi \sim 1$ (composition of hydrogen and oxygen only). The column density grows from top to bottom, while the rest of the parameters are held constant (i.e., constant $\xi$ and composition). Note that in such a setup, the ionization profile is independent of where the matter is truncated. On the left, the three panels show the ionization fraction of oxygen $O^{5+}$ to $O^{8+}$. At the illuminated face of the matter, the $O^{8+}$ fraction is set by the ionization-recombination equilibrium. This value then decreases on a scale $\tabs\simeq1$. On the right, the corresponding transmitted spectra are shown, with the incident spectra given for reference. In the top simulation, there is nearly no absorption, the middle one corresponds to marginal absorption, and the bottom one shows significant absorption. The gray dashed line marks the absorption by neutral matter of the same composition and column density.}
    \label{fig:ion_frac}
\end{figure}

For this simple case, there are two options \taya{for the ionization state at the illuminated side of the matter}:\\ 
(i) The ionizing radiation cannot fully ionize a significant fraction of the oxygen atoms at $r_\text{in}$, i.e., $1-\yi \ll 1$, where $\yi \equiv y(r_\text{in})$. In this case, above $\sim 1\rm~ keV$, the neutral cross-section is a good approximation for the absorption cross-section at the illuminated side of the slab, while further in, the ionization level is even lower. Thus, regardless of the column density, the absorption cross-section is approximately $\sigma_{O7\to 8}$, and $\tabs\simeq \fo\Sigma\sigma_{O7\to 8}$. Note that below 1 keV, the lower ionization states of oxygen may shape the absorption, and their ionization states are not specified in this calculation.
(ii) At the illuminated side of the slab, a non-negligible fraction of the oxygen is fully ionized: i.e., $1-\yi \sim 1$. In this case, the ionization will remain as in the illuminated side over a column density of  $\Delta\Sigma\sim \frac{1}{\yi\fo\sigma_{O7\to 8}}$, and we have three scenarios (demonstrated in figure \ref{fig:ion_frac}). (a) If  $ \yi\fo\Sigma\sigma_{O7\to 8}  \ll 1$, then the ionizing X-ray photons are not absorbed significantly during their propagation, $y(r) \approx \yi$, so $\tabs \ll 1$ at all frequencies and locations in the slab, resulting in no significant absorption (example at the upper panel of figure \ref{fig:ion_frac}). (b) If $\yi \fo\Sigma\sigma_{O7\to 8} \gg 1$, the column density is high enough for the ionization degree, $y$, to decrease significantly within the slab, so for a significant fraction of the column density $1-y \ll 1$, ionizing photons are absorbed and the neutral cross-section becomes a good approximation (lower panel in figure \ref{fig:ion_frac}). (c) The last option is that $\yi\fo\Sigma\sigma_{O7\to 8}  \sim 1$, in which case detailed modeling of the ionization is required to find the X-ray absorption (middle panel in figure \ref{fig:ion_frac}). 

Next, we derive the ionization state at the illuminated side of the slab, $\yi$. Assuming that the oxygen is composed only of $O^{7+}$ and $O^{8+}$, the equation for ionization-recombination equilibrium is given by: 
\begin{equation}\label{eq:ion-rec} n_{e}n_{O^{8+}}\alpha_{O8\to 7}\left(T\right)=n_{O^{7+}}\frac{1}{4\pi r^{2}}\dot{N}_{\gamma,i}\cdot\left\langle \sigma_{O7\to 8}\right\rangle \end{equation} 
where $\alpha_{O8\to7}\left(T=10^5 {\rm~K}\right)\simeq 9\cdot10^{-12}{\rm~cm^3/s}$ is the B-type recombination coefficient, which depends on the temperature roughly as $T^{-0.5}-T^{-0.75}$ \citep{Storey1995,Badnell2006}, $n_e$ is the electron number density, which is approximately equal to the hydrogen number density, $\dot{N}_{\gamma,i}$ is the number flux of ionizing photons (with energy above the $O^{7+}$ ionization threshold - $\varepsilon_i\sim 0.5$ keV), and $\left\langle \sigma_{O7\to 8}\right\rangle$ is the mean ionization cross-section weighted by the number of incident photons at every energy. As we limit our discussion to $L_\nu\propto\nu^{\alpha}$ with $\alpha \leq 0$, both the flux of ionizing photons and the photons absorbed are dominated by photons with $h\nu\simeq\varepsilon_i$, $\dot{N}_{\gamma,i}\cdot\left\langle \sigma_{O7\to 8}\right\rangle \approx \dot{N}_\gamma(\varepsilon_i)\sigma_{0,O7\to 8}$ where $\dot{N}_\gamma(\varepsilon_i)\approx L_\nu(\varepsilon_i)$ is the number of photons with energy comparable to $\varepsilon_i$, and $\sigma_{0,O7\to 8}\equiv\sigma_{O7\to 8}(\varepsilon_i)$ is the cross-section at the ionization threshold. For $O^{7+}$, the ionization threshold is $\varepsilon_i\sim 0.5~{\rm keV}$\footnote{The $O^{7+}$ ionization threshold is 0.87 keV. However, for simplicity, and in order to later include the ionization thresholds of nearby ions, \taya{such as the iron L-shell,} we use $\varepsilon_i=0.5$ keV. \taya{This introduces a negligible increase in the number of ionizing photons considered of ($\frac{\dot{N}_\gamma(0.5-0.87\rm ~keV)}{\dot{N}_\gamma(h\nu >0.87\rm ~keV)}$), a change that is absorbed in the calibration constants when we calibrate our model to numerical simulations.}}, and the ionization cross-section is: $\sigma_{0,O7\to 8}(\varepsilon_i)\simeq \sigma_{O7\to 8}(h\nu=0.87 {\rm keV})\simeq 9.8\cdot10^{-20} {\rm~cm^2}$ \citep{Verner1996}.

We now define the ionization parameter\footnote{Note that the ionization parameter can be defined in various forms, sometimes not including $c$, or not specifying the integration limits. Here we use the dimensionless quantity and specify the limits of the integral to start from the ionization threshold.} - the ratio between the local number density of the ionizing photons and the electron number density:

\begin{equation}\label{eq:xi} 
\begin{aligned} \xi &\equiv \frac{1}{4\pi r^2 n_e c}\intop_{\varepsilon_i}^{\infty}\frac{L_{\nu}}{h\nu}d\nu \approx \frac{\dot{N}(\varepsilon_i) }{4\pi r^2 n_e c} \\&\approx 0.17 \frac{\nu L_\nu(1~ {\rm~keV})}{10^{40} {\rm~erg~s^{-1}}}\left(\frac{\Sigma}{10^{24} {\rm~cm^{-2}}}\right)^{-1} \left(\frac{r}{10^{14} {\rm~cm}}\right)^{-1}~. \end{aligned} 
\end{equation}

In the last term, we provided an approximate expression using canonical values that are appropriate for SN CSM interaction, assuming the CSM has a wind density profile. We can use equations \eqref{eq:ion-rec} \& \eqref{eq:xi} to express the ratio of fully ionized to non-fully ionized oxygen at the inner boundary: 

\begin{equation} 
\frac{1-\yi}{\yi} = \frac{n_{O^{8+}}}{n_O-n_{O^{8+}}} \approx \frac{n_{O^{8+}}}{n_{O^{7+}}}=\frac{\sigma_{0,O7\to 8} c}{\alpha_{O8\to 7}\left(T\right)}\xi. 
\end{equation}

This immediately gives us the first criterion for when the neutral absorption cross-section is a good approximation. If $1-y \ll 1$, or equivalently, $\xi\ll \xi_c\sim\frac{\alpha_{O8\to 7}\left(T\right)}{\sigma_{0,O7\to 8}(\varepsilon_i) \cdot c}\simeq 0.003$, we expect the neutral cross-section to be an excellent approximation, at least above $\sim 1\rm~keV$, beyond the  $O^{7+}$ ionization edge, where contributions from lower ionization states are negligible. Thus, we define $\xi_c$ as the critical value of $\xi$ below which neutral absorption always holds (above $1~\rm keV$), and in \S\ref{sec:numerical_thin} we numerically calibrate its value for realistic compositions and find $\xi_c=0.015$. If $\xi \gg \xi_c$, we expect a significant fraction of the oxygen at $r_\text{in}$ to be fully ionized, and the effective absorption cross-section at the illuminated side of the matter to be $\yi \sigma_{0,O7\to 8}(\varepsilon_i)$. At any frequency, the radiation conditions in the matter change on a scale of $\tabs(\nu)\simeq 1$. Therefore, while $\tabs(h\nu\approx\varepsilon_i)\lesssim1$, the X-ray will pass unabsorbed. Thus, when $y_{in} \lesssim 1$, the X-ray will be unabsorbed if:

\begin{equation} 
\yi\fo\Sigma\sigma_{0,O7\to 8}\simeq \fo \Sigma \frac{\alpha_{O8\to 7}\left(T\right)}{c \xi} \lesssim 1. 
\end{equation} 

The last dependence we need to consider is the dependence of $\alpha_{O8\to 7}$ on temperature and the dependence of temperature on the matter conditions. For this, we make the following simplifying assumption: we assume the temperature scales like the energy deposited per electron and is a monotonic function of $\xi$. We also assume $\alpha_{O8\to 7}$ is a monotonic function of $T$, and that the resulting $\alpha_{O8\to 7}(\xi)$ can be expressed as a power-law in $\xi$ over a limited range, so that $\alpha_{O8\to 7} \propto \xi^{1-\beta}$, where $\beta$ is a constant. Under this assumption, and the assumption that in gas with realistic composition, the fraction of absorbers (analogous to $\fo$ in the oxygen only toy model) is proportional to the gas metallicity, we define the following parameter, which corresponds to $\tabs(h\nu\simeq1\rm ~keV)^{-1}$ for $\tabs(h\nu\simeq1\rm ~keV)<1$: 
\begin{equation}\label{eq:W_def} 
W\equiv \frac{\xi^{\beta}}{\sigma_{eff}\Sigma \left(\frac{Z}{Z_{\odot}}\right)} ~.
\end{equation} 
Here $\sigma_{eff}$ is a calibration constant, $\sigma_{eff}\sim\sigma_{neutral,\odot}\left(h\nu=0.5 ~{\rm keV}\right)/\xi_c^{\beta}$. In the following section, we numerically calibrate $\sigma_{eff}$ and $\beta$ so that $W\simeq1$ corresponds to marginal absorption and separates the regimes of no absorption at $W\gg1$ and nearly neutral absorption at $W\ll1$. $\beta$ is chosen so that the transition between the two regimes will occur over the smallest possible range in $W$.
\taya{\subsubsection{Additional remarks}
\begin{itemize}
    \item  In our discussion above, we limited the spectra to cases where $\dot{N}_{\gamma,i}$ is dominated by photons with energy $\sim \epsilon_i$. Our results can be extended to SEDs where the number flux of ionizing photons is dominated by higher energies, i.e., $\alpha>0$. However, then the product $\dot{N}_{\gamma,i} \cdot \left\langle \sigma_{O7\to 8}\right\rangle$ needs to be handled more carefully, as it is spectrum dependent. While we do not address such SEDs in this work, we expect that generalizing our result to specific SEDs with a rising number flux is straightforward.
    \item In the regime we are considering, the free electron number is typically dominated by ionized hydrogen and helium and is therefore independent of the ionization state. Furthermore, even at high metallicities, the electron number density changes little across the ionization states where X-ray photoabsorption varies most dramatically. This is because the dominant absorption features (K-edges) disappear when the innermost electrons are ionized, but at these high ionization states, most electrons are already free. For example, the transition from $O^{7+}$ to $O^{8+}$ eliminates the oxygen K-edge but only increases the electron contribution from oxygen by $12.5\%$ (1 out of 8 electrons)
    \item A consequence of the previous remark is that $\xi$ does not vary with metallicity.
\end{itemize}}

\subsection{Numerical validation}\label{sec:numerical_thin}

To validate and calibrate our analytical results, we use \textsc{Cloudy} \citep{CLOUDY2023}. We run simulations in spherical symmetry\footnote{Not to be confused with the "sphere" setting in \textsc{Cloudy}, which is set to off}, with a luminosity source at the origin, 
\begin{equation}
L_\nu\propto\nu^{\alpha}\exp\left(-\frac{h\nu}{\kB T_0}\right), h\nu\ge 1{\rm~eV}
\end{equation}
with $\alpha=-2,-1,-0.5,0$,  a cutoff temperature of $T_0=3\cdot10^7-3\cdot10^9 \rm~K$. Motivated by the density profile expected for stellar winds, we test density profiles of $n = n_0 (r/r_\text{in})^{-2}$ at $r>r_\text{in}$, as well as constant density profiles. The density range in our simulations is $n\simeq 10^7-10^{15} {\rm~cm^{-3}}$, consistent with the regime in which three-body recombination is negligible (as previously discussed in footnote \ref{fn:three-body}). Our model should also be valid for lower densities and perhaps slightly higher densities. In our tests, we find that a thin slab ($\Delta r\lesssim r$) of constant density and a wind profile with the same density at $r_\text{in}$ and the same column density yield almost identical results. Therefore, we expect our results to be valid for any profile where most of the column density is near the inner radius.

\textsc{Cloudy} does not treat Thomson scattering correctly. In fact, we find that for fully ionized matter, the incident radiation attenuation due to scattering is $\sim \exp(-\tau_T)$ rather than the $\sim\tau_T^{-1}$ attenuation predicted by steady state diffusion. This over-prediction becomes important at $\tau_T\simeq0.3$; therefore, we limit our simulations to lower column densities. We vary the metallicity in the range $z=0.01-50 z_{\odot}$. This allows us to explore higher column densities of metals without increasing the electron scattering optical depth, thus keeping the simulations within the Thomson-thin regime. For simplicity, all simulations are run with the "no scattering" option, but include Compton heating. 

\taya{Another limitation of \textsc{Cloudy} is its treatment of lines. The line width in \textsc{cloudy} is set artificially by the grid, and the escape-probability formalism is a good approximation only when lines are extremely optically thin, or extremely optically thick. While this work is not focused on lines, in principle, as the lines are often important for the cooling, one could expect these issues to alter the temperature estimations, and thus the ionization state. The main dependence of the ionization state on temperature is through the recombination coefficient in the relevant temperature range for line emission, $\alpha_B\propto T^{-0.5}, ~10^4{\rm ~K}\le T\le 10^6{\rm ~K}$. This means that the temperature must change by more than an order of magnitude to change the ionization state by more than a factor of a few. We do not expect lines becoming optically thick to change the temperature enough to be a significant effect for the following reasons. Firstly, a significant effect on the temperature is expected only if all the lines that allow efficient cooling are optically thick, and this typically is not the case. Secondly, if a line becoming optically thick would lead to heating the matter significantly, the heated matter would then decrease the optical depth of the line, both through the $\alpha_B$ dependence and by increased collisional ionization, and the line would become optically thin again.  This expectation is supported by the work of \cite{Dumont2000}. Using their Monte-Carlo photoionization code, \textsc{TITAN}, they run several simulations varying only the column density by two orders of magnitude, and find that the temperature profile changes by $\lesssim30\%$ with the changes to the column density, and that the temperature estimate by \textsc{Cloudy} is a good approximation. We conclude that the temperature estimates are likely good and depend primarily on the ionization parameter, not on the exact optical depth of the lines, though this claim should be tested once a relevant code becomes available. }

Using these simulations, we validate our model assumptions (see Appendix \ref{App:Numerical validation}) and calibrate our model. We calibrate $\xi_c$ by plotting the absorption cross-section at 0.5 keV (at the illuminated side of the cloud) as a function of $\xi$, and we find that, up to $\xi\simeq0.015$, it is roughly constant; at higher values of $\xi$, it starts to decline steeply. From this, we read $\xi_c\simeq0.015$.  

Next, we calibrate $\sigma_{eff}$ and $\beta$ by treating our simulation output as a mock observation and fitting it with the same incident spectrum, attenuated by neutral absorption. By this process, we can match each of our simulations with a neutral equivalent column density - $\Sigfit$. Since most X-ray telescopes are only sensitive above 0.2 keV, and there is insignificant absorption above 10 keV, we only fit the range of 0.2-10 keV. In addition, as observations have a limited signal-to-noise ratio, we fit only the part of the transmitted spectrum that is brighter than $1/30$ of its peak in $\nu L_\nu$.

With this analysis, we can compare $\Sigfit$ to the actual column density, $\Sigma$, as a function of $W$ to find where the neutral approximation is reasonable, where there is significant absorption but the neutral approximation fails, and where there is no significant absorption. We calibrate $\sigma_{eff}$ to provide a transition between neutral-like absorption at $W\ll1$ and no absorption at $W\gg1$ and $\beta$ to minimize the range of $W$ over which this transition occurs. We find, $\beta=1$ and $\sigma_{eff}=6\cdot 10^{-25} {\rm~cm^{2}}$, that the calibrated expression for $W$ is: 
\begin{equation}\label{eq:W_val}
\begin{aligned}
        W &\approx 1.7 \xi \left(\frac{\Sigma}{10^{24} {\rm cm^{-2}}}\right)^{-1} \left(\frac{Z}{Z_\odot}\right)^{-1}\\
    &\approx 0.3 \frac{\nu L_\nu(1~\rm keV)}{10^{40}~\rm erg/s}\left(\frac{r}{10^{14}}\right)^{-3}\left(\frac{n}{10^{10}}\right)^{-2}\left(\frac{Z}{Z_\odot}\right)^{-1}~, 
\end{aligned}
\end{equation}
\taya{And here, too, the second expression is an approximation for a wind profile.} 

\begin{figure}
    \centering
    \includegraphics[width=\columnwidth]{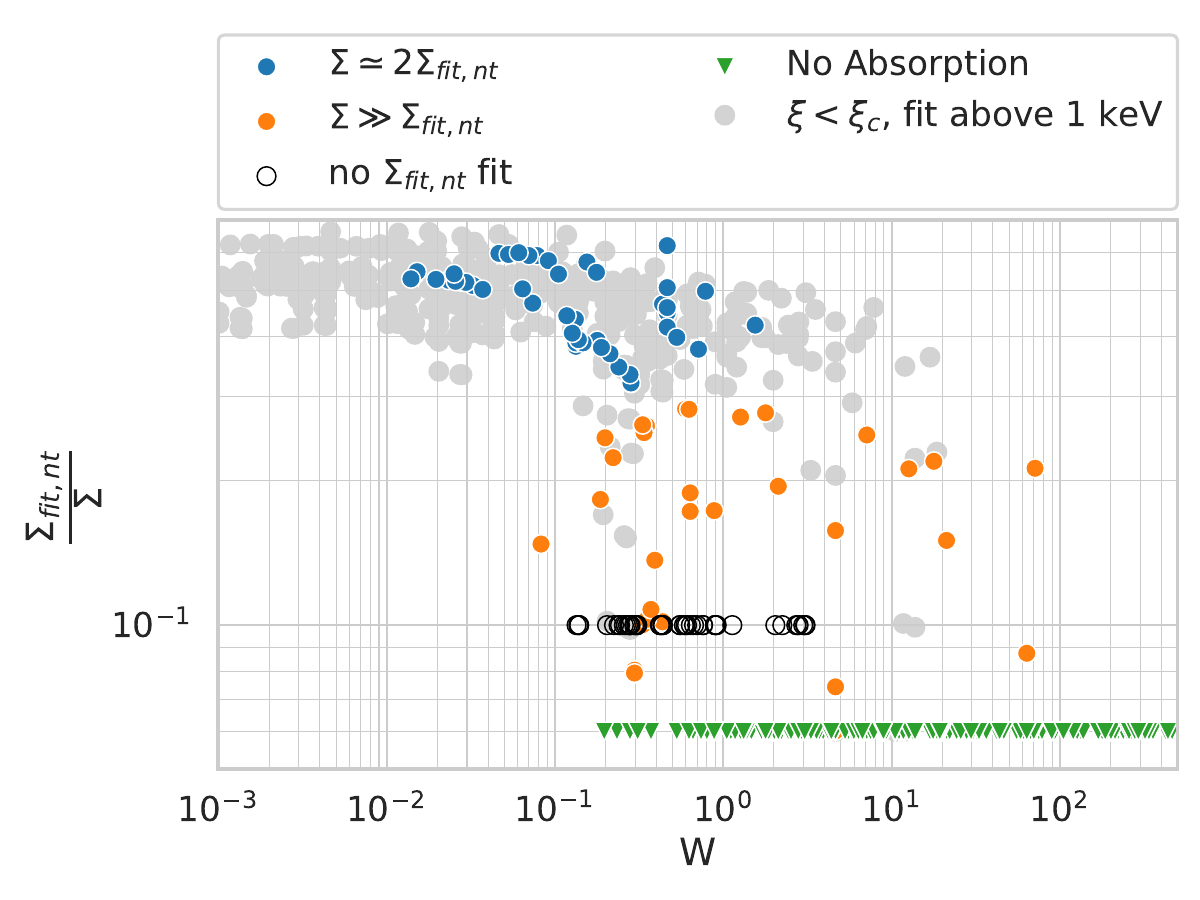}
    \caption{For each of the simulations we ran, we plot the ratio between the best-fit column density obtained under the assumption of fully neutral gas, $\Sigfit$, and the actual column density $\Sigma$, as a function of $W$ (Eq. \ref{eq:W_val}). Every simulation corresponds to a single dot.  {\it Grey dots}: simulations in which $\xi\le\xi_c$. For these, the neutral column density is fit only above 1 keV, because below it, non-neutral features appear. An example of such a spectrum can be seen in the second panel of figure \ref{fig:example}. Cases in which $\xi\ge \xi_c$ are marked in various colors, according to the fit to neutral column density. {\it Blue dots}: cases in which the neutral column density that fits the spectrum is approximately half the actual column density. An example is given in the top panel of figure \ref{fig:example}. {\it Orange dots}: cases in which a good fit can be found for absorption by neutral column density, but the fit results in a value much lower than the actual column density (example in the third panel of figure \ref{fig:example}). {\it Green triangles}: simulations for which there is no significant absorption, and only an upper limit on the neutral column density can be found. For readability, all these simulations are plotted at the same value, although each results in a different upper limit (example in the bottom panel of figure \ref{fig:example}). {\it Black hollow circles}: simulations for which no neutral column density can provide a good fit; the value of $\Sigfit/\Sigma$ is meaningless and is arbitrarily set to 0.1 in the figure for clarity. An example of such a case is given in the fourth panel of figure \ref{fig:example}.}
    \label{fig:Sigma_fit}
\end{figure}
The ratio $\dfrac{\Sigfit}{\Sigma}$ as a function of $W$ is shown in figure \ref{fig:Sigma_fit}. It shows that the absorption is approximately neutral at $W\ll1$ or when $\xi\le \xi_c$, and that there is negligible absorption at $W\gg1$, as expected. We also mark two intermediate cases: one in which the neutral absorption fits the spectrum but has a column density much lower than the actual column density, and the other in which the absorption features cannot be fitted well by neutral absorption at any column density. 

\begin{figure}
    \centering
    \includegraphics[width=\columnwidth]{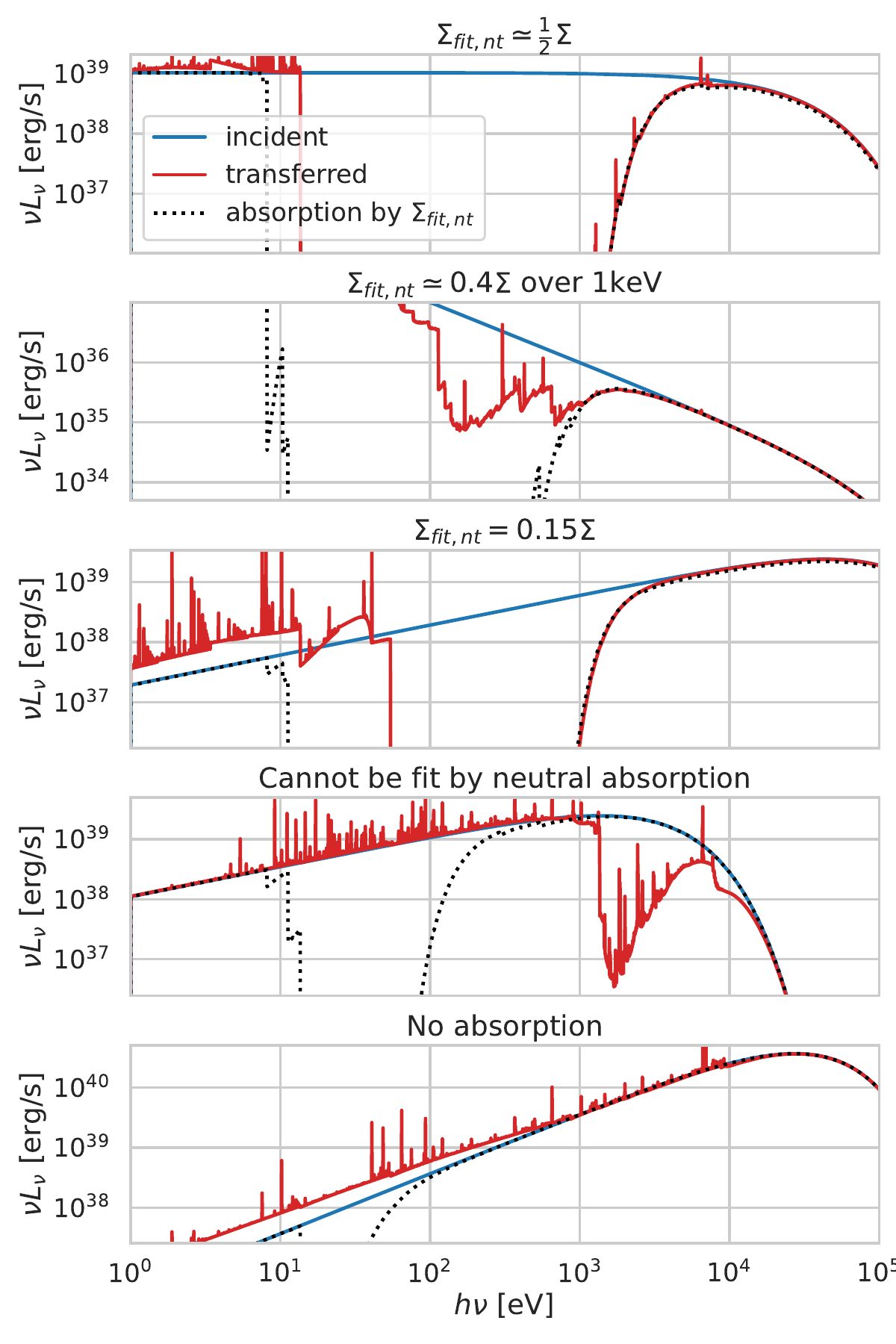}
    \caption{Example spectra from our simulations, showing the incident spectrum, the transmitted spectrum, and the best-fit neutral absorption spectrum.  
    Top panel: approximately neutral column density, corresponding to orange dots in figure \ref{fig:Sigma_fit}. Second panel: spectrum with $\xi<\xi_c$, where the neutral cross-section is a good approximation only above 1 keV, corresponding to the gray markers in figure \ref{fig:Sigma_fit}. Third and fourth panels: spectra from the transition region; the third panel shows a spectrum where the best-fit cross-section is significantly below the actual cross-section, and the fourth panel shows a spectrum where absorption features are not well-described by a neutral fit. In figure \ref{fig:Sigma_fit}, these are marked in orange and hollow circles, accordingly. Fifth panel: negligible absorption, where all elements up to oxygen are fully ionized. Such simulations are marked in green triangles in figure \ref{fig:Sigma_fit}.
    }
    \label{fig:example}
\end{figure}

The four cases we find are demonstrated in figure \ref{fig:example} and differ in the modeling they require. The first case (figure \ref{fig:example} upper panel) is when the neutral absorption is a good approximation, but the inferred column density is approximately half of the actual column density. In this case, as predicted, most of the column density has oxygen that is not fully ionized, either because there are not enough ionizing photons for any of the oxygen to be fully ionized ($\xi<\xi_c$) or because the column density is high enough for the ionization state to change considerably ($W\lesssim0.1$). The opacity is not equivalent to a fully neutral column since lighter elements are ionized and lower ionization levels are not occupied; these make up roughly half the absorption at $\sim$ keV. \footnote{In the limit of $\xi\to0$, if there are no soft photons, the matter will be totally neutral and  $\Sigfit=\Sigma$. Our work is limited to the regime in which hydrogen is fully ionized, and the transition to completely neutral matter is not explored here.} $\Sigfit\simeq \frac{1}{2}\Sigma$. A caveat to this relation is that in very low metallicities ($Z\ll 0.1Z_\odot$), when there is a significant column of partially ionized oxygen (and other heavy metals) but helium is fully ionized, then $\Sigfit\ll \frac{1}{2}\Sigma$ because, in such metallicities, helium dominates the X-ray opacity in fully neutral gas. Note that while $W\lesssim0.1$ ensures that the column density is high enough for the ionization state to change considerably within the matter, $\xi<\xi_c$ only promises that none of the Oxygen will be fully ionized. If, in addition, the column density is low ($W\ge 0.1$), non-neutral features will be visible below 1 keV due to other ions (example in figure \ref{fig:example}, second panel).

The second case is when a good fit to the data can be found with a neutral column density, but the resulting value is much smaller than the actual column density due to significant ionization (example in figure \ref{fig:example}, third panel). This happens when the column density is high enough that the ionization state farther into the medium is lower than at the illuminated side, but the ionized part of the medium makes up a considerable fraction of the column density.  
The third case is when neutral absorption cannot describe the absorption features (example in figure \ref{fig:example}, fourth panel). In this case, the value of $\Sigfit/\Sigma$ is meaningless, since the fit is not good. For readability, in figure \ref{fig:Sigma_fit}, we arbitrarily set it to 0.1.  
The fourth case is when all the elements up to oxygen have a high enough ionization fraction that there is negligible or no absorption (example in figure \ref{fig:example}, lower panel). This regime assumes that lighter elements, including hydrogen and helium, are efficiently ionized. Here too, if the metallicity is low  ($Z\lesssim 0.1Z_{\odot}$), the ionization state of helium can change the expected absorption state. At higher metallicities, the secondary electrons from the ionization of metals heat the matter, and helium is collisionally ionized. At lower metallicities, the matter is not sufficiently heated for that, and singly ionized helium can cause significant absorption in the X-ray range; therefore, the ionization state of Helium must be evaluated separately.

Our recommendations for applying these results to observations are summarized in Table \ref{tab:W}.
\begin{table*}
\begin{threeparttable}
    \centering
\begin{tabular}{|c|c|c|}
\hline 
$W<0.1$ or $\xi \ll \xi_c$ & $0.1\lesssim W \lesssim 1$ or & $W \gg 1$ and $\xi \gg \xi_c$ \tabularnewline
&$\xi\simeq\xi_c$ and $W>1$&
\tabularnewline 
\hline 
\hline 
 $\Sigma \simeq (2-3) \Sigfit^{*}$ & Numerical modeling is required & No absorption$^{*}$\tabularnewline
 (neutral fit is good at least above 1 keV$^{\dagger}$) & possibly $\Sigfit\ll\Sigma$, or no good fit & for $0.2 \lesssim h\nu \lesssim 10$ keV \tabularnewline
\hline
%\tabularnewline
% $\xi<\xi_c$ and $W>0.1$, possibly non-neutral features at $0.2-1$ keV & In some cases, $\Sigfit\ll\Sigma$ & for $Z\le0.1$ and $\alpha\ge-0.5$,\tabularnewline
% For $Z/Z_\odot\lesssim0.1$, ionized Helium can reduce $\Sigfit$ & or no good neutral fit &  possibly He absorption up to  $W\sim100$\tabularnewline
% \hline 
%\tabularnewline \tabularnewline
\multicolumn{1}{c}{}
\\
\hline
\multicolumn{3}{|c|}{Relevant parameter space}\tabularnewline
\hline 
\hline 
Spectrum & Column density & density and density profile \tabularnewline
\hline 
$L_{\nu}\propto\nu^{\alpha}\exp\left(-\frac{h\nu}{kT_0}\right)$,
$\alpha\le0$ (tested for $-2 \le \alpha \le 0$), & $1.2\cdot10^{21}{\rm ~cm^{-2}}\left(\frac{Z}{Z_{\odot}}\right)^{-1}\le\Sigma\le1.5\cdot10^{24}{\rm ~cm^{-2}}$ & $n=n_{0}\left(\frac{r}{r_\text{in}}\right)^{-k},k>1$ (or $\Delta r\lesssim r_\text{in}$), \tabularnewline
$T_0\le4\cdot10^{9}$ K,
No induced Compton (see Appendix \ref{sec:Induced Compton}). &  & 
and $n_{0}\le10^{15}{\rm~cm^{-3}}$. \tabularnewline
&&\tabularnewline
\hline 
\end{tabular}
\begin{tablenotes}
\footnotesize
\item[$^{*}$] For $Z\lesssim0.1Z_{\odot}$, in neutral matter, Helium dominates the absorption at $\sim$ keV energies. Since helium ionizing sources are unaccounted for, fully ionized helium can reduce absorption when neutral absorption is expected. On the other hand, we find that for low metallicities and hard spectra ($\alpha\simeq -0.5-0$), the helium is not always fully ionized, even when the heavier elements are, and can cause significant absorption. 
\item{$^{\dagger}$} $\xi \ll \xi _c$ corresponds to oxygen that is not fully ionized. However, that only ensures approximately neutral absorption above 1 keV. If the column density is too low for the ionization state to change significantly in the matter, i.e, $W\gtrsim 0.1$, there will be non-neutral absorption features in the 0.3-1 keV range. When $W\lesssim0.1$,  the column density is high enough for the ionization fraction to change significantly, and cause neutral absorption also of lower ionization states.
\end{tablenotes}
     \caption{Top: absorption properties depending on $W$ (equation \ref{eq:W_val}) and $\xi$ (equation \ref{eq:xi}) relative to $\xi_c$ ($\xi_c=0.015$). Bottom: parameter space where our model is relevant.}\label{tab:W}
\end{threeparttable}
\end{table*}

\taya{\subsection{Implications for emission and absorption lines}\label{sec:lines}
Our model naturally has implications for emission and absorption lines. $W\gg1$ corresponds to a state in which the matter is ionized enough (up to the iron K-shell) that the absorption column density in the X-ray is below unity. This is also expected to significantly reduce the absorption lines originating from the elements up to iron (excluding the iron K-shell), corresponding to inhibiting all absorption lines below $\sim 5 \rm ~keV$. Emission lines may be inhibited for a different reason. The steep dependence of the recombination coefficients on the temperature means that emission lines are inhibited when the matter temperature is high. While a full investigation of this effect is beyond the scope of this paper, the temperature dependence on $\xi$, presented in figure \ref{fig:T(xi)}, can be used to estimate the recombination coefficient for specific lines, as needed. 
Lastly, the prominent iron lines, which originate from the K-shell electrons, are impacted. The 6.4 keV line, which requires an L-shell electron, is not expected when $W\gg1$. The 6.7 and 7 keV lines, which originate from He-like and H-like iron, respectively, are expected when the gas temperature is in the several keV to tens of keV range.}

\section{Thomson thick media}\label{sec: Thomson thick}
To derive properties of the transferred spectrum in the Thomson thick case, we must generalize the absorption model described in \S\ref{sec: Thomson Thin} and consider several additional effects: Thomson scattering, Compton heating of the matter by the X-ray photons, Comptonization, and Compton degradation of the radiation field.\footnote{One additional effect that becomes important is absorption by iron. For matter at solar metallicity, the absorption cross-section of iron at $\sim$10 keV is of the order of the Thomson cross-section, which is why it was insignificant in the Thomson thin case. This means that in the Compton thick case, even when we predict that the X-ray ionizes its way out, there may be a signature of absorption in iron lines in the X-rays and an absorption edge from k-shell iron above $\sim$9 keV. Detailed modeling of this is beyond the scope of this work.} The relative importance of these various effects depends on how many times a photon scatters through the medium, and that, in turn, depends on the boundary condition experienced by photons scattered back towards the source. We treat two cases for this boundary condition. The first is a reflective boundary, in which the photons scattered back towards the source are reflected; this boundary is a good description, for example, of a point source surrounded by a thick shell, with very low density in between. The second is a reprocessing boundary, in which the photons scattered back towards the source reach a dense, cool medium, in which photons below $\sim$ 10 keV are efficiently reprocessed to lower energies. This boundary is appropriate, for example, for some regimes of interacting SNe. As we cannot simulate the Thomson-thick regime, our predictions for this case are rougher and should be verified numerically in detail once a code becomes available. Specifically, our derivation does not treat geometrical effects, which may alter our results by factors of order unity.

We build upon the results from the previous section; thus, we adopt similar assumptions. Here too, we assume $n\lesssim10^{15}\rm ~cm^{-3}$ and that the column density is dominated by the matter in the vicinity of the inner radius, $r_{in}$, with the hydrogen fully ionized. We also require some conditions specific to this regime; we assume here that the time it takes for radiation to diffuse through the medium in the Thomson thick regime is short relative to the time it takes for the state of the system to change, and we consider Thomson optical depths of $1\lesssim\tT\lesssim 100$. However, the results below, when used with $\tT=1$, should also be valid for $\tT\lesssim1$, and we expect that most of our results here will also hold also for $\tT\gtrsim 100$. 

Note that for sufficiently large optical depths, matter and radiation will always approach thermal equilibrium, and the transmitted spectrum will approach black-body radiation. The details of the transition to this state are beyond the scope of this work. However, we can outline some cases in which this is expected. For example, if significant absorption is expected (the conditions for this are described below in Tables \ref{tab:reflective} and \ref{tab:reprossesing}) and $\tT\gtrsim 10$, the combination of Compton degradation and photo-absorption leads to a reprocessing of all the energy in X-rays and possibly to thermal equilibrium. If the emission is expected to be transferred unabsorbed, free-free and Compton effects need to be considered to determine when free-free absorption becomes significant enough for the matter to reach thermal equilibrium.  

Below, we start by discussing the effect of electron scattering and generalizing the absorption model to the Thomson thick case. We follow by discussing the additional changes to the spectrum due to Compton effects. Finally, we consider the boundary conditions for photons scattered back towards the source and treat each of the two considered boundaries in detail.

\subsection{Thomson scattering}
When photons scatter back and forth through a section of the medium, they contribute to two competing effects that both shape the ionization of the medium. The first is that they increase the local ratio of ionizing photons to electrons, thereby increasing the ionization of the medium. The second is that every photon passes through the same medium several times, increasing the effective column density it experiences and increasing the chance of being absorbed. If a photon scatters through a section of the medium $k$ times, it effectively increases $\Sigma$  by a factor of $k$, and, to first order, also increases the local value of $\xi$ by a factor of $k$ (in some cases, there are higher-order effects due to the dependence of the gas temperature on $\xi$, which we treat later.).

If $\xi>\xi_c$ and the value of $W$ determines whether the radiation will pass unabsorbed, then since $W\propto\frac{\xi}{\Sigma}$, these two effects cancel out. Thus, as long as the net flux through the medium is not changed, the X-ray absorption (or lack thereof) is independent of the scattering properties of the medium. We verify this using a 1D Monte-Carlo simulation described in Appendix \ref{App:Monte-Carlo}. Hence, in the Thomson thick case, we use the same definitions of $\xi$ (Eq. \ref{eq:xi}) and $W$ (Eq. \ref{eq:W_val}) as in the Thomson thin case, with only minor modifications at each of the two boundary conditions considered here.

\subsection{Compton effects}
The mean net energy gain of a photon with energy $E_\gamma \ll m_e c^2$ due to (spontaneous) Compton collisions with thermal electrons at a temperature of $\kB T_e \ll m_e c^2$ is \citep{RybickiLightman1986}:
\begin{equation}\label{eq: comp_scat}
    \Delta E_\gamma = \frac{4\kB T_e E_\gamma-E_\gamma^2}{m_e c^2}.
\end{equation}
Photons with energy below $4\kB T_e$ will be Comptonized, while photons with energy above it will lose energy to Compton degradation.  

\subsubsection{Compton degradation}
A photon with energy $E_\gamma$ diffusing through matter in which $E_\gamma \gg 4 \kB T_e$ loses approximately $\frac{\Delta E_\gamma}{E_\gamma}\simeq-\frac{E}{m_e c^2}$ on every collision, and after $N$ collisions, all photons with an initial energy $E>\max(\frac{m_e c^2}{N},4 \kB T_e)$ are lowered to a temperature of $\simeq\max(\frac{m_e c^2}{N},4 \kB T_e)$, setting a new cutoff in the spectrum. 

\subsubsection{Comptonization}
Comptonization can up-scatter photons from lower energies into the keV range, increasing the effective value of $\xi$, and thus changing the local ionization equilibrium. 
A photon with energy $E_\gamma$ scattering through matter with a temperature $\kB T_e\gg E_\gamma$ gains, on average, $\frac{\Delta E}{E}=\frac{4 \kB T_e}{m_e c^2}$ per scatter. Thus, the number of scatters needed to bring a photon with energy $h\nu$ to the keV range (so that it will be included in $\xi$):
\begin{equation}
    N_{scat} = \frac{m_ec^2}{4 \kB T_e}\cdot \log\left(\frac{{\rm 1~keV}}{h\nu}\right)
\end{equation}
 or, equivalently, if a photon scatters $N_{scat}\gg \frac{m_ec^2}{4k_BT_e}$ times in the hot region, the number of photons up-scattered into keV energies will grow by $\simeq \exp(-\alpha N_{scat}\cdot\frac{4 \kB T_e}{m_e c^2})$, where $\alpha$ is the spectral index.
This effectively increases the value of $\xi$ by that factor as well.

\subsubsection{Compton heating of electrons}
In these scatterings, the photons can also heat or cool the electrons. The electron temperature at which the mean change in photon energy due to a collision with thermal electrons vanishes is called the Compton temperature and is given by\footnote{Note that here we focus only on cases where induced Compton heating is unimportant. We shortly discuss cases where it does in \S\ref{sec:Induced Compton}.}: 
\begin{equation}\label{eq:Tc}
    \TC\equiv \frac{h}{4 \kB}\frac{\intop \nu L_\nu d\nu }{L}.
\end{equation}
If the Compton temperature is high enough (usually $\TC\gtrsim10^7$ K is sufficient), the temperature will approach the Compton temperature when $\xi$ is high. This will be shown in the following section.

\subsection{The electron temperature}
In the Thomson thin regime, we considered the impact of electron temperature on the observed spectrum only via its effect on the recombination coefficient. This effect was calibrated to the regime in which absorption is significant, i.e., $W\lesssim 0.1$, which corresponds to  $\xi\lesssim0.1$ for Thomson thin media. In the Thomson thick case, the high column densities make it possible to attain absorption also with much higher values of $\xi$, for which the temperature is no longer set by ionization equilibrium but by Compton heating. Therefore, the value of $\alpha_B$ must be corrected for the matter temperature in these cases. In addition, a high electron temperature may affect the observed broadband X-ray spectrum in the Thomson thick regime in several additional ways. First, as discussed above, $T_e$ sets a lower limit to Compton degradation. Second, the free-free emission of the medium may be detectable (i.e., comparable to or brighter than the incident radiation at some frequencies). Third, Comptonization of lower energy photons can increase the number density of photons above $\sim 0.5$ keV, effectively changing the value of $\xi$. Finally, one may expect that in some scenarios, collisional ionization may cause the gas to become fully ionized when photoionization alone is insufficient to do so. However, we show in Appendix \ref{App:collisional_ion} that this is not the case.

Therefore, in Thomson thick medium, one must estimate $T_e$ (which may be a function of $\tau_T$, along the medium). The electron temperature is determined by line cooling, free-free cooling, photoelectric heating, and Compton heating and cooling. In the general case, the equilibrium temperature depends on the ionization states of the different ions and therefore should be solved numerically. Since the temperature along the medium depends on many parameters, we cannot provide a simple formula here to determine its value. Instead, we provide the results of a large set of numerical simulations that enable the reader to obtain a rough estimate of $T_e(\tau)$, where needed, for a given set of parameters, by approximating the effective value of $\xi$ in the region of interest. We use these to identify the parts of phase space in which finding the temperature is unnecessary and the regions where the matter is at the Compton temperature. 

\begin{figure}
    \centering
    \includegraphics[width=0.9\columnwidth]{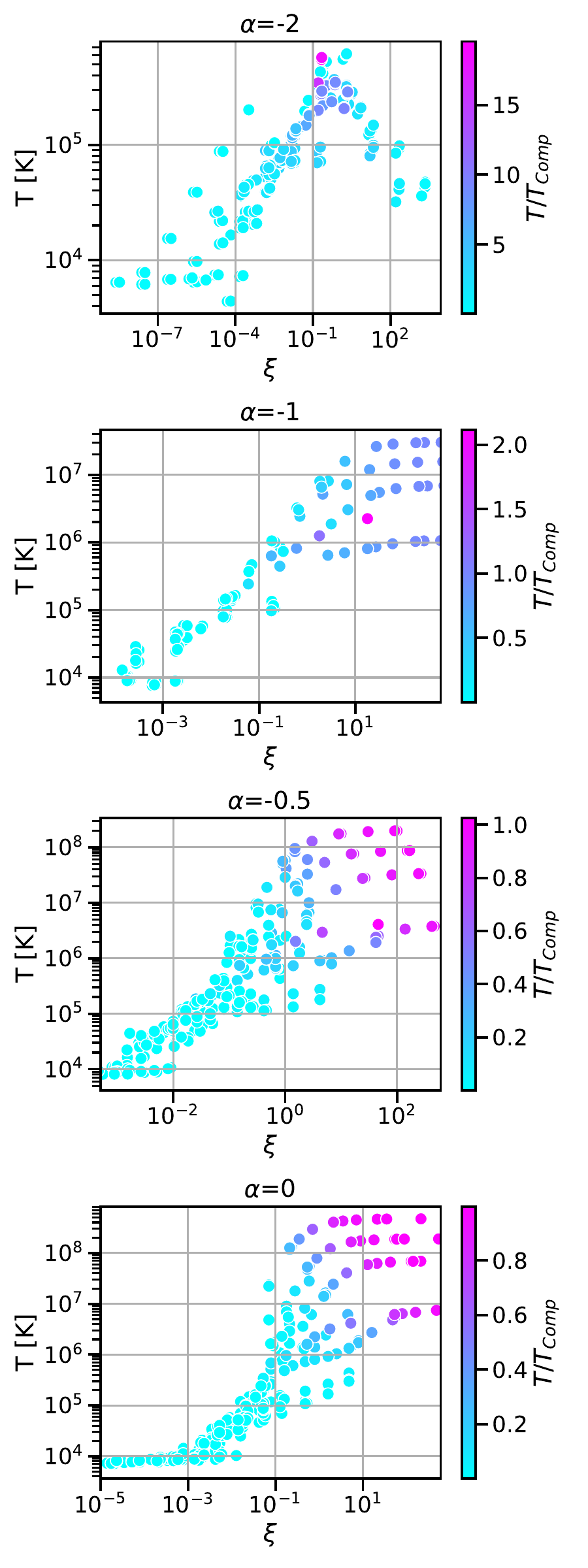}
    \caption{The matter temperature at the illuminated side of the medium is given as a function of $\xi$, for each of our simulations (all Thomson thin). The color bar shows the temperature relative to the Compton temperature (equation \ref{eq:Tc}). Note that for Thomson thick matter, scattering increases the effective local value of $\xi$ by a factor of $\sim \tT$, and to read the expected temperature from this graph, one must compare the X-axis of this plot to the value of $\tT\cdot \xi$ \taya{for a reflective boundary and to $\xi \frac{\tT}{\tT(r_{\rm in})}$ for a reprocessing boundary, assuming negligible absorption.}}
    \label{fig:T(xi)}
\end{figure}

In figure \ref{fig:T(xi)}, we present numerical results for the temperature at the illuminated side of the matter as a function of $\xi$ for several power-law spectra, \taya{based on the Thomson thin regime, for which we were able to run simulations. To use this graph to estimate the temperature profile in the medium, one must estimate $\xi(\tT)$ and then read the temperature from this graph.} Each panel focuses on a different value of $\alpha$ and shows the values for various simulations ranging in cutoff temperature and metallicity. For $\alpha=-2$, the temperature is always set by a balance of photoionization heating and line cooling. As the Compton temperature is low and Compton cooling is subdominant, the temperature can easily exceed the Compton temperature. Still, it is always far below $10^7$ K. For $\alpha=-1,-0.5,0$, this description is valid for low values of $\xi$, but as $\xi$ grows, the temperature rises to $T\gtrsim  10^7~\rm K$, and the temperature is set by the balance between Compton heating and free-free cooling. For larger values of $\xi$, the matter reaches the Compton temperature. 
Using these results, for every value of $\alpha$ tested here, we define a value of $\xi$ above which the matter temperature is above $10^7$ K, \taya{i.e., line cooling is not efficient} and Compton heating is significant. At $\xi\gg\xiC$, the matter can be assumed to be at the Compton temperature. Note that if the Compton temperature is below $10^7$ K for any reason, Compton heating can be neglected. We list the values of $\xiC$ in table \ref{tab:xi_Comp}.
\begin{table}
\centering
    \begin{tabular}{|c|c|c|c|c|}
        \hline
         $\alpha$ & -2  & -1  &-0.5  & 0  \\
         \hline
         $\xiC$  & N/A  & 0.1 & 0.3  & 1 \\
         \hline
    \end{tabular}
    \caption{The value of $\xiC(\alpha)$. For $\xi>\xiC$ Compton heating becomes important.}
    \label{tab:xi_Comp}
\end{table}

Using the various components outlined above, we can now consider the boundary condition and outline the emergent spectrum. 

\subsection{Reflective boundary}
Consider a reflective boundary condition such that photons scattered back towards the source are perfectly reflected. The emergent spectra depend on whether the matter is heated to the Compton temperature and on the ionization state. Using the derivations and arguments from the previous subsections, we can derive the temperature and ionization state of the matter and predict the absorption and additional features of the spectrum. Below, we describe the various regimes and then summarize the expected absorption state in Table \ref{tab:reflective}.

The temperature and ionization state of the medium depend on the local ionizing photon density. We start by finding how many times photons scatter through a given part of the medium\footnote{As we are primarily interested in heating, and if the Compton temperature is high, the photons heating the matter are the high-energy ones ($>10$ keV), we can neglect absorption in this case.}. For that, we will use $\tT$, the Thomson optical depth to the observer from a given point in the matter, to mark the location within the medium, and we denote $\tau_{T,in} \equiv \tau_T(r_\text{in})$ as the Thomson optical depth from the inner radius to the observer. Using random walk arguments and the requirement for a steady state, it is easy to show that a photon scatters through the matter an average of $\sim \tTin^2$ times before escaping and passes through a point with $\tT>1$, about $\tT$ times in the process, increasing the local number density of ionizing photons by the same amount. Thus, the photon density at any point with $\tT>1$ is proportional to $\tT$, and the actual ionization parameter in the Thomson thick part of the medium is $\simeq\xi\tT$ (where $\xi$ is the ionization parameter we defined for the Thomson thin medium, in Eq. \ref{eq:xi}, and $\tT\ge1$). Thus, any point in which $\xi\tT\gg \xiC$, the gas is heated to the Compton temperature, and at any point where $\xi\tT \ll \xiC$, the gas temperature is below $10^7$ K. Using this, we divide our phase space into three cases according to the temperature of the matter.

First, consider the case in which the entire medium has a temperature $T\lesssim 10^7$ K, such that Compton heating is negligible. This occurs either when the incident spectrum makes such heating impossible, i.e., $\alpha<-2$ so $\TC<10^7$ K, or if even the matter most easily heated, the matter at $\tTin$, has a temperature $<10^7$ K. That is, $\xi\tTin\ll\xiC$. 
If the temperature in the entire medium is below $10^7$ K, the conditions for the incident spectrum to be absorbed are similar to those derived in the Thomson thin case. Namely, if $\xi<\xi_c$, then at least where $\tT\le1$, the oxygen is not fully ionized, and we expect neutral-like absorption in that section of the matter. If $\xi>\xi_c$, the absorption depends on $W$ (as in Eq. \ref{eq:W_val}, with no modifications) and follows the same criteria described in the Thomson thin case; if $W\ll0.1$ there is significant absorption, if $W\gg1$ there is no absorption, and in between, numerical modeling is required.  Finally, regardless of the value of $\xi$, in this regime, Compton degradation limits the energy of the emerging photons to $m_e c^2/\tTin^2$. Note that in this regime, if $\tTin\gtrsim7$ and there is significant absorption, the X-ray may be entirely absorbed since Compton degradation reduces all photon energies to below 10 keV, where they can be efficiently reprocessed. If $\tT\gtrsim30$ and the matter is sufficiently cold (below a few times $10^6$ K), Compton degradation may set the cutoff to below 0.5 keV, inhibiting X-ray transmission altogether.

Next, we can consider a case in which the entire medium is Compton heated. This occurs when $\xi \gg \xiC$, and the Compton temperature is sufficiently high  ($\TC\gtrsim10^7$ K).
Before considering the ionization state, we need to take into account two corrections due to the temperature. First, the increased temperature causes $\alpha_B$ to decrease. Second, in some cases, Comptonization of lower energy photons increases the number of photons in the $\sim$ keV range. These two effects can be incorporated into the ionization modeling by correcting $W$. We thus define:
\begin{equation}\label{eq:tildeW}
    \tilde{W}=W\cdot\left(\frac{\TC}{10^{6}{\rm ~K}}\right)^{0.7}\cdot\begin{cases}
    \exp\left(-\alpha\tTin^{2}\cdot\frac{4\kB \TC}{m_{e}c^{2}}\right), & \tTin^2 \gg \frac{m_e c^2}{4 \kB \TC}\\
1,&\tTin^2 \lesssim \frac{m_e c^2}{4 \kB \TC}.
\end{cases}
\end{equation}
where the first term takes into account the dependence of $\alpha_B$ on $T$ and the second term accounts for the potential increase in the number of ionizing photons due to Comptonization. With these corrections, when $\tilde{W}\ll0.1$ a significant part of the medium is not fully ionized, we expect significant absorption, while if $\tilde{W}\gg1$, then absorption is negligible. In between, modeling is required to determine the state of the matter. Note that as $\tilde{W}\ge W$, if $W\ge 1$, that is sufficient for there to be no absorption. Finally, in this regime, Compton degradation is limited to $4\kB \TC$. In addition, when calculating the transmitted spectrum, one should take into account the effect of Comptonization and free-free emission of the hot medium. 

Lastly, we consider the intermediate case in which only part of the matter is Compton heated. A full solution of this case requires numerical modeling since the energy transferred between the hot and cold regions by Compton scattering, simultaneously with its effect on the ionization states, is difficult to model. While we do not carry out a full treatment here, the transmitted spectrum can be modeled for several limiting cases in this regime.
First, if we assume that Compton heating is negligible, the radiation can still ionize its way through; surely it can do so when part of the matter is Compton heated. Hence, in this case, we expect the incident radiation to pass through unabsorbed. Second, if there is significant absorption even when we assume that the entire matter is Compton heated, then certainly there is significant absorption when only part of the matter is Compton heated, too. Additional information is obtained from the fact that in a partial heating state, the matter at $\tT\le1$, which has the lowest ionization parameter, is not heated. Thus, we can check if the matter at $\tT\sim 1$ is fully ionized by verifying if $\xi$, after being corrected for Comptonization, is still lower than $\xi_c$. In which case, we also expect significant absorption.
We thus define: 
\begin{equation}
    \tilde{\xi} = \xi \cdot\begin{cases}
    \exp\left(-\alpha\tTin^{2}\cdot\frac{4\kB \TC}{m_{e}c^{2}}\right), & \tTin^2 \gg \frac{m_e c^2}{4 \kB \TC}\\
1,&\tTin^2 \ll \frac{m_e c^2}{4 \kB \TC}
\end{cases} 
\end{equation}
and conclude that if $\tilde{\xi}\le \xi_c$, there is significant absorption. We note that even in these limiting cases, Compton degradation, Comptonization, and free-free emission may still impact the transmitted spectrum in a way that we cannot estimate without having a full solution for the gas temperature profile.

We summarize the expected absorption in Table \ref{tab:reflective}. 
\begin{table*}
\centering
\textbf{Reflective boundary} \\[0.3em]
\begin{threeparttable}
    \centering
\begin{tabular}{|c|c|c|c|}
\hline 
$\xi<\frac{\xiC}{\tTin}$, or $\alpha<-2$, or $\TC<10^{7}K$ & \mbox{$\xi\ll\xi_{c}$} or \mbox{$W<0.1$}  & \mbox{$0.1\lesssim W\lesssim1$}  & \mbox{$W\gg1$}\tabularnewline
\cline{2-4} \cline{3-4} \cline{4-4} 
Matter is cold ($\lesssim 10^7$ K) & Significant absorption up to \textasciitilde 10 keV$^{*}$ & Numerical modeling is required  & No absorption$^{*\dagger}$\tabularnewline
\hline 
\hline 
$\frac{\xiC}{\tTin}\lesssim\xi\lesssim\xiC$ and $\TC\gtrsim10^{7}$ K & \mbox{$\tilde{W}\ll0.1$ or \mbox{$\tilde{\xi}\ll\xi_{c}$} } & \mbox{$\tilde{W}\gtrsim0.1$} and $\tilde{\xi}>\xi_c$ and \mbox{$W\lesssim1$}  & \mbox{$W\gg1$}\tabularnewline
\cline{2-4} \cline{3-4} \cline{4-4} 
Part of the matter is Compton heated & Significant absorption up to $\sim$10 keV$^{*\ddagger}$  & Numerical modeling is required$^{\ddagger}$  & No absorption$^{*\dagger\ddagger}$\tabularnewline
\hline 
\hline 
$\xi\gg\xiC$ and $\TC\gtrsim10^{7}K$ & \mbox{$\tilde{W}\ll0.1$}  & $0.1\lesssim\tilde{W}\lesssim1$  & \mbox{$\tilde{W}\gg1$}\tabularnewline
\cline{2-4} \cline{3-4} \cline{4-4} 
Matter fully Compton heated & Significant absorption up to \textasciitilde 10 keV$^{*\ddagger}$ & Numerical modeling is required$^\ddagger$  & No absorption$^{*\dagger\ddagger}$ \tabularnewline
\hline 
\end{tabular}
\begin{tablenotes}
\footnotesize
\item[$^{*}$] For $Z\lesssim0.1Z_{\odot}$, in neutral matter, Helium dominates the absorption at $\sim$ keV energies. Since helium ionizing sources are unaccounted for, fully ionized helium can reduce absorption when neutral absorption is expected. On the other hand, we find that for low metallicities and hard spectra ($\alpha\simeq -0.5-0$), the helium is not always fully ionized, even when the heavier elements are, and can cause significant absorption. 
\item[$^{\dagger}$] Iron k-shell electron ionization is not fully modeled and may cause absorption above $\sim 9$ keV. 
\item[$^\ddagger$] Spectrum may show signs of Comptonization and free-free emission. 
\end{tablenotes}
\caption{We summarize in which cases absorption is expected for Thomson thick media with a reflective boundary. The prediction for absorption (or lack therof) are given in terms of$ \xi$ (equation \ref{eq:xi}), $\xi_c=0.015$, $W$ (equation \ref{eq:W_val}), $\xiC$ (table \ref{tab:xi_Comp}), and $\tilde{W}$ (equation \ref{eq:tildeW}). Note that in addition to absorption, the spectrum is also shaped by Compton degradation, Comptonization, and free-free emission.\\
\textbf{To use this table}, first calculate $\xi$ and $\TC$ for your incident spectrum, and check the first column of the table to see if the matter is cold  ($<10^7$ K), partially Compton-heated, or fully Compton-heated. For your relevant rows, the next columns contain criteria for $W$ or $\tilde{W}$ to check whether you expect significant absorption.} \label{tab:reflective}
\end{threeparttable}
\end{table*}

\subsection{Reprocessing boundary}
Assume that on the other side of the source, there is cold dense matter. Photons below 10 keV will be reprocessed for the following reason: photons scattered into the reprocessing region are absorbed if, at their energy, the cross section for absorption is higher than the Thomson cross section, or reflected if it is smaller.\footnote{At these energies, the cross section for bound Compton scattering is comparable to Compton scattering on free electrons, so this claim is independent of whether the hydrogen is ionized in the reprocessing region.} For solar metallicity, the neutral absorption cross section per hydrogen atom is larger than the Thomson cross section below $\sim$ 10 keV. This means that for photons below 10 keV, the boundary condition is absorbing, while for photons above 10 keV, the boundary condition is approximately reflective\footnote{The high-energy photons may lose energy in the reprocessing region due to Compton degradation, \taya{and a fraction of them will be absorbed, e.g., by photoionizing iron}.}. To predict the transferred spectrum, we must model the temperature and ionization conditions in every part of the medium between the source and the observer. The ionization is dominated by the softer, $<10$ keV photon population, while the temperature may be dominated by \taya{either of the populations}. We therefore need to understand both populations. 

We start by describing the behavior of these two populations in the limit where there is no absorption within the medium, neglecting corrections due to the effect of the temperature of the medium on the radiation field. All of these will be accounted for later on. 
Considering photons below 10 keV, by random walk arguments, the probability of a photon escaping to the observer without crossing over to the reprocessing region is $ \tTin^{-1}$. Thus, when there is no absorption in the medium between the source and the observer, the luminosity of photons below 10 keV that reach the observer in the reprocessing boundary case is reduced by a factor of $\tTin$ compared to the Thomson thin and reflective boundary cases. 
To find the number density of the photons that have not been reprocessed in the Thomson thick part of the medium, we recall that, as there are no sources or sinks, steady state dictates a constant net flux, which implies, due to diffusion, a number density that is proportional to $\tT$. This means that at every point along the Thomson thick medium, the number density of the photons below 10 keV is $\frac{\tT}{\tTin}$ times that of the Thomson thin case (assuming no absorption).  

Photons above 10 keV behave as in the reflective boundary case and scatter $\sim \tTin^2$ times as they diffuse out, enhancing the local photon density by a factor of $\tT$ relative to the Thomson thin case.
Thus, the spectrum at any point in the medium will have a "jump" of a factor $\tTin$ at around 10 keV. This means the spectrum will no longer be strictly decreasing, as assumed in the Thomson thin and reflective boundary cases with no absorption. To correct the use of $\xi$ for this case, we must consider the two uses we have for $\xi$. The first is to quantify the ionization state. For this purpose, we do not care about the "extra" photons above 10 keV for the following reason: Comparing the ionization cross-section of photons at 1 keV to the ionization cross-section of $\tTin$ times more photons at 10 keV, we must consider that the ionization cross-section decreases as $\sim E_\gamma^{-3}$. For $\tau_T\lesssim100$, the contribution of ionizing photons above 10 keV can be neglected relative to lower energy photons, while for larger $\tT$, Compton degradation in the inner cold reprocessing boundary will limit the photon's energy to far below 10 keV. Thus, for this purpose, we can assume the entire spectrum is reprocessed at the inner boundary. The other use for $\xi$ is to estimate whether the radiation can bring the medium to the Compton temperature. For this use, the high-energy photons may be the dominant ones. Hence, we define:
\begin{equation}\label{eq:xi10}
    \xiten =\frac{\dot{N}_\gamma(h\nu>10 {\rm ~keV})}{4\pi r^2nc}.
\end{equation}
With this definition in hand, we can derive the matter temperature. We are interested in the broadband properties of the transmitted spectrum, especially in determining which part of the X-ray spectrum of the source is absorbed by the intervening medium and which part is transmitted. Thus, we consider only bound-free absorption and ignore line emission and absorption. Using \textsc{Cloudy} simulations, we find that for spectra truncated below 10 keV, $\xiten\gtrsim\xiCten\equiv 0.6$ is sufficient to heat the matter to the Compton temperature. $\xiCten$ is independent of $\alpha$, since the spectrum in this case extends over a single order of magnitude in energy, or less. We conclude that for spectra with $\alpha>-2$, and a high energy cutoff above 10 keV, the matter at some optical depth $\tT$ is hot if either $\frac{\tT}{\tTin}\xi\ge\xiC$ or $\tT \xiten\ge\xiCten$. Note that for $\alpha\ge-2$, while the reprocessing slightly decreases the Compton temperature, by shifting energy to lower frequencies, this effect is negligible. For $\alpha<-2$, reprocessing can significantly decrease the Compton temperature, but the Compton temperature, in this case, is close to the low-energy cutoff and is not a relevant parameter anyway. Using the new criteria for Compton heating and recalling that the photon number density below 10 keV is decreased because of the reprocessing, we can adapt the results from the reflective boundary case. The description of the expected absorption is summarized in Table \ref{tab:reprossesing}. The main differences between the reprocessing and reflective boundary cases are that $W$ is replaced by $\frac{W}{\tTin}$ to account for the photons being reprocessed, and that the heating can depend also on the high-energy part of the spectrum and is expressed in terms of $\xiten$. 

There are two more effects that require consideration due to the new shape of the spectrum in this case, though they turn out not to be important. The first is that the additional heating leads to more efficient collisional ionization by thermal electrons. We explore this in Appendix \ref{App:collisional_ion}, and find that for $\tTin\lesssim40$, photoionization is the dominant ionization source. The second is that photons above 10 keV can be Compton degraded into the keV range if $\tT\ge 10$, thereby increasing the number of ionizing photons. However, once they are degraded to below 10 keV, if they pass through the boundary, they will be reprocessed. The probability of them scattering enough times to be degraded down to the keV range without being reprocessed is $\sim \tTin^{-1}$, thus they can at most increase the ionizing photon density by a factor of order unity. 

\begin{table*}
\centering
\textbf{Reprocessing boundary} \\[0.3em]
\begin{threeparttable}
    \centering
\begin{tabular}{|P{4cm}|c|P{4cm}|c|}
\hline 
\mbox{$\xiten\ll\xiCten$} and \mbox{$\xi/\tTin\ll\xiC$} or \mbox{$\TC<10^{7}{\rm~K}$} or \mbox{$\alpha\le-2$} & \mbox{$\frac{\xi}{\tTin}\ll\xi_{c}$} or \mbox{$\frac{W}{\tTin}<0.1$} & \mbox{$0.1\le\frac{W}{\tTin} \le1$} & \mbox{$\frac{W}{\tTin}\gg1$}\tabularnewline
\cline{2-4} \cline{3-4} \cline{4-4} 
Matter is cold & Significant absorption below \textasciitilde 10 keV$^*$ & Numerical modeling required & No absorption$^{*\dagger}$\tabularnewline
\hline 
\hline 
$\xiCten/\tTin\le\xiten\le\xiCten$ or $\xiC\le\xi\le\tTin\xiC$
and $\TC>10^{7}{\rm~K}$ & $\frac{\tilde{W}}{\tau_{in}}\ll0.1$ or $\tilde{\xi}/\tTin\ll\xi_{c}$ & $\tilde{\xi}/\tTin\gtrsim \xi_c$ and $0.1\lesssim W/\tTin$ and $\tilde{W}/\tTin\lesssim1$ & $W/\tau_{in}\gg1$\tabularnewline
\cline{2-4} \cline{3-4} \cline{4-4} 
Part of the matter is Compton heated & Significant absorption below \textasciitilde 10 keV$^{*\ddagger}$ & Numerical modeling required$^\ddagger$ & No absorption$^{*\dagger\ddagger}$\tabularnewline
\hline 
\hline 
\mbox{$\xiten>\xiCten$} or \mbox{$\xi/\tTin>\xiC$} and \mbox{$\TC>10^{7}K$} & $\tilde{W}/\tau_{in}\ll0.1$ &$0.1\lesssim\tilde{W}/\tau_{in}\lesssim1$ &  $\tilde{W}/\tau_{in}\gg1$\tabularnewline
\cline{2-4} \cline{3-4} \cline{4-4} 
Matter fully Compton heated & Significant absorption below \textasciitilde 10 keV$^{*\ddagger}$. & Numerical modeling required$^\ddagger$ & No absorption$^{*\dagger\ddagger}$\tabularnewline
\hline 
\end{tabular}
\begin{tablenotes}
\footnotesize
\item[$^{*}$] For $Z\lesssim0.1Z_{\odot}$, in neutral matter, Helium dominates the absorption at $\sim$ keV energies. Since helium ionizing sources are unaccounted for, fully ionized helium can reduce absorption when neutral absorption is expected. On the other hand, we find that for low metallicities and hard spectra ($\alpha\simeq -0.5-0$), the helium is not always fully ionized, even when the heavier elements are, and can cause significant absorption. 
\item[$^{\dagger}$] Emission below $\sim$ 10 keV is suppressed by a factor of $\tTin$ relative to emission at higher energies. Iron k-shell electron ionization is not fully modeled and may cause absorption above $\sim 9$ keV. 
\item[$^\ddagger$] Spectrum may show signs of Comptonization and free-free emission. 
\end{tablenotes}
\caption{We summarize in which cases absorption is expected for Thomson thick media with a reprocessing boundary. The prediction for absorption (or lack thereof) are given in terms of the $\tTin$ - optical depth to the source, $\xi$ (equation \ref{eq:xi}), $\xi_c=0.015$, $\xiten$ (equation \ref{eq:xi10}) $W$ (equation \ref{eq:W_val}), $\xiC$ (table \ref{tab:xi_Comp}), $\xiCten=0.6$, and $\tilde{W}$ (equation \ref{eq:tildeW}). Note that in addition to absorption, the spectrum is also shaped by Compton degradation, Comptonization, and free-free emission. This table is valid in the same range of parameters listed in Table \ref{tab:W}, except the column density of relevance (expressed in terms of Thomson optical depth) is $1\le \tT\lesssim 40$.\\
\textbf{To use this table}, first calculate $\xi$ and $\TC$ for your incident spectrum, and check the first column of the table to see if the matter is cold  ($<10^7$ K), partially Compton-heated, or fully Compton-heated. For your relevant rows, the next columns contain criteria for $W$ or $\tilde{W}$ to check whether you expect significant absorption.  
}
    \label{tab:reprossesing}
\end{threeparttable}
\end{table*}

In addition to the absorption, the spectrum is also shaped by Compton degradation, the Comptonization of low-energy photons in the hot region, and the free-free emission of the hot region.
\section{Examples From Supernovae}\label{sec:examples}
\taya{Below, we discuss X-ray observations of two supernovae in the context of our model. The aim of these is to provide an example of applying our model and drawing conclusions from it, rather than testing a specific model.}
\subsection{SN\,2023ixf (Thomson thin medium)}
We consider the X-ray observations of SN\,2023ixf between 4 and 60 days. All these epochs have NuSTAR spectra, as well as lower-energy spectra from XRT/CXO/XMM. \cite{Nayana2025} fit the X-ray spectrum using the assumption of neutral absorption, and measure the column density in this manner between days 4.4 and 58.4. They estimate the column density in a second way, via the total X-ray luminosity. They find that the two measurements agree during the first epoch. But, at later times, the X-ray luminosity method results in column densities that are higher by about an order of magnitude than the columns measured from absorption. \taya{They suggest a change to ionization level, clumps, and CSM asymmetry as three ways to reconcile this discrepancy.} 

Here, we use our model to check whether the assumption of a neutral medium is self-consistent with the observations. 
We use the luminosities and column densities measured by \cite{Nayana2025} (their table 1), and the assumption that the velocity of the shock (which is the X-ray source) is roughly constant,  $10^9 {\rm cm/sec}$, as measured from optical spectra \citep{Zimmerman2024,Dickinson2025}, to estimate the values of  $\xi$ and $W$ (Eqs. \ref{eq:xi} and \ref{eq:W_val}) at different times. These are presented in Table \ref{tab:ixf}. As we are working under the assumption of approximately neutral absorption, we use $\Sigma\simeq 2-3 \Sigma_{nt,abs}$ in calculating $W$.

\begin{table*}
    \centering
\begin{tabular}{|c|c|c|c|c|}
\hline 
time (days) & $\xi$ (from absorption) & $W$ (from absorption) & $\xi$ (emission) & $W$(emission)\tabularnewline
\hline 
\hline 
4.4 & 0.01 & 0.02-0.04 & 0.008 & 0.03\tabularnewline
\hline 
11.5 & 0.04 & 0.6-1 & 0.004 & 0.02\tabularnewline
\hline 
21.4 & 0.03 & 1.2-1.9 & 0.002 & 0.03\tabularnewline
\hline 
30.5 & 0.04 & 3-5 & 0.001 & 0.03\tabularnewline
\hline 
58.4 & 0.01 & 2-3 & 0.0007 & 0.02\tabularnewline
\hline 
\end{tabular}
    \caption{$\xi$ and $W$ calculated based on the X-ray observations of SN\,2023ixf reported in \protect\cite{Nayana2025}. We report $W$ and $\xi$ based both on the column density they infer from absorption \taya{(where we use $\Sigma=2-3\Sigma_{nt,abs}$ to translate neutral equivalent column density to column density)}, and on the density value they infer from the X-ray emission. In the first epoch, we conclude that the assumption of neutral absorption is consistent. At later epochs, since the observations show absorption, but the value of $W$ derived based on the absorption is inconsistent with neutral absorption, we conclude that these epochs fall within the transition region, where ionization is significant and the system needs to be modeled carefully.}
    \label{tab:ixf}
\end{table*}

We find that the consistency of the first epoch with neutral absorption is marginal as $\xi \approx \xi_c$, and  $W \approx 0.1$. Therefore, it is possible that the actual column density is 2 to 3 times larger than measured under the neutral assumption, namely, $\Sigma\simeq 6-9\cdot 10^{23}{\rm ~cm^{-2}}$, but to be sure, detailed modeling is recommended. \taya{The values from emission suggest a similar conclusion for this epoch.}

From the second epoch onward, we find that using the column density \taya{as measured based on absorption}, the values of $W$ obtained are \taya{either in the transition regime, which requires modeling, or} in the regime where no absorption is expected, \taya{while using the column densities inferred from emission, provides values of $\xi<\xi_c$ for these epochs, which should have implied neutral absorption}. We conclude that the assumption of an approximately neutral column density is inconsistent. Therefore, the actual column in these epochs is higher than measured under the neutral assumption, and given that absorption features are observed, the system is in a part of the phase space where numerical modeling is required. These results explain, at least in part, the discrepancy between the column densities found by the two methods from the second epoch onward. Our analysis suggests that the ionization state of the matter should be investigated more closely, but does not rule out that global asymmetry or local inhomogeneities may also play a part in shaping the observations of SN\,2023ixf \citep{Nayana2025}.

\subsection{SN\,2008D (Thomson thick medium)}
The early X-ray observations of SN\,2008D are interpreted as a shock breakout from either a star or a wind, \taya{or emission from a relativistic jet} \citep{Soderberg2008,Chevalier2008,Mazzali2008,Modjaz2009,Svirski2014b}. Here, we demonstrate the use of our model to check to consistency of X-rays diffusing through a thick wind and remaining unabsorbed. We adopt the radius and Thomson optical depth assessed by \cite{Svirski2014b}, \taya{the only model requiring $\tau_T>1$, and hence the one for which the lack of absorption is most constraining.} Namely, $r=6\cdot10^{11}\rm cm$, and $\tTin\approx4$, and the range of X-ray luminosities observed by \cite{Modjaz2009} in the first 300 seconds, $L_x(0.2-10{\rm~keV})\simeq 3-30 \cdot 10^{42} {\rm erg/s}$. 
Using these, we calculate $\xi$ and $W$. We use $\dot{N}_\gamma\simeq\frac{L}{0.5 \rm ~keV}$ and approximate the density profile as a wind profile, i.e., $n_e= \frac{\Sigma}{r_{in}}=10^{13} {\rm~ cm^{-3}}$. Using these, we find $\xi\simeq 3\cdot 10^3-3\cdot 10^4$, and $W\simeq 850-8500$. Since $W\gg1$, and $\tilde{W}\ge W$, we do not need to check the temperature and calculate $\tilde{W}$, and the radiation should pass unabsorbed, regardless of the matter temperature. In addition, for $\tTin=4$, we would not expect to see signatures of Compton degradation or Comptonization, so we conclude this model is consistent with little to no X-ray absorption, \taya{implied from the low neutral equivalent column density of $N_{H}\approx 5\cdot 10^{21}{\rm ~cm^{-2}}$ [$<1.5\cdot 10^{21}{\rm ~cm^{-2}}$] found by \citep{Modjaz2009} when fitting to a power-law [black body] model.}

\section{Summary}\label{sec:summary}
We study the problem of an X-ray source obscured by an intervening medium, for both Thomson thin and Thomson thick media. For Thomson thin medium ($\tT\ll1$), our main result is a simple set of criteria that points to the appropriate way that a transmitted spectrum should be analyzed, based on the source and intervening medium properties. These criteria are calibrated and validated using numerical radiative transfer modeling with \textsc{Cloudy}  and are summarized in Table \ref{tab:W}, alongside the parameter space in which they are valid. For the Thomson thick case ($\tT\gtrsim1$), we consider a reflective and reprocessing inner boundary condition. We find that in the Thomson thick regime, if the electron temperature in the irradiated medium is above $10^7$ K,  it has a significant effect on the transmitted spectrum. Therefore, we provide a set of criteria to determine if the entire intervening medium is cold, hot, or partially hot and partially cold. For each of these regimes, we provide a set of tools for deriving the general absorption properties of the transmitted spectrum. We also discuss other processes that can affect the observed spectrum, such as Compton degradation and free-free emission. The expected absorption is summarized in Tables \ref{tab:reflective} and \ref{tab:reprossesing} for the reflecting and reprocessing cases, respectively. 

The base of our absorption model is an analytical model for the Thomson thin regime, validated and calibrated with \textsc{Cloudy} simulations. Since oxygen is the main absorber in the $\sim$ keV energy range, we analytically derive a toy model for absorption by matter composed of hydrogen and oxygen. We then validate and calibrate it using \textsc{Cloudy}, considering metallicities of $0.01-50Z_\odot$. We find a simple criterion for when the X-ray is able to ionize its way out of the medium unabsorbed and when the approximation of absorption by a neutral column density is valid (up to minor calibration). Between these two regimes, there is a regime for which the X-ray absorption must be modeled carefully, either because the neutral absorption column that fits can be much smaller than the actual column density or because the absorption features are very different from those formed by a neutral absorber. 

For the Thomson thick case, we do not have a simulation we can use. Instead, we analytically extend the Thomson thin model and consider, in addition, the effects of electron scattering and Compton degradation. We focus on two cases for the boundary conditions experienced by photons reflected back towards the source. A reflective boundary, in which photons scattered back towards the source are reflected, and a reprocessing boundary, in which photons scattered back reach neutral material that can efficiently reprocess the X-rays below $\sim 10$ keV and reflect photons above 10 keV. For both cases, we consider the effects of Compton heating and Comptonization on the ionization balance and absorption, and we derive when significant absorption is expected, when the radiation can photoionize its way out, and when a more detailed analysis is required. 
 Since the transmitted spectrum depends on many parameters, such as the Compton temperature of the radiation, which can vary along the medium (e.g., by Compton degradation), the resulting transmitted spectrum needs to be calculated carefully for each set of parameters. Here, we provide the tools to derive the gas temperature and ionization, so the reader can calculate the transmitted spectrum for any system of interest (unless it falls within the regime that requires numerical modeling).  Further work is needed to verify and calibrate the Thomson thick case once numerical simulations become available.  

\section*{Acknowledgments}
We thank Hagai Netzer, Jonathan Stern and Raffaella Margutti for the useful discussions. 
This research was partially
supported by a consolidator ERC grant 818899 (JetNS) and an
ISF grant (1995/21).

%%%%%%%%%%%%%%%%%%%%%%%%%%%%%%%%%%%%%%%%%%%%%%%%%%
\section*{Data Availability}
The data underlying this article will be shared upon reasonable request to the corresponding author.



\bibliographystyle{mnras}
\bibliography{refs} % if your bibtex file is called example.bib

@ARTICLE{Wilms2000,
       author = {{Wilms}, J. and {Allen}, A. and {McCray}, R.},
        title = "{On the Absorption of X-Rays in the Interstellar Medium}",
      journal = {\apj},
         year = 2000,
        month = oct,
       volume = {542},
       number = {2},
        pages = {914-924},
          doi = {10.1086/317016},
archivePrefix = {arXiv},
       eprint = {astro-ph/0008425},
 primaryClass = {astro-ph},
       adsurl = {https://ui.adsabs.harvard.edu/abs/2000ApJ...542..914W}
}

@ARTICLE{CLOUDY2023,
       author = {{Chatzikos}, M. and {Bianchi}, S. and {Camilloni}, F. and {Chakraborty}, P. and {Gunasekera}, C.~M. and {Guzm{\'a}n}, F. and {Milby}, J.~S. and {Sarkar}, A. and {Shaw}, G. and {van Hoof}, P.~A.~M. and {Ferland}, G.~J.},
        title = "{The 2023 Release of Cloudy}",
      journal = {\rmxaa},
         year = 2023,
        month = oct,
       volume = {59},
        pages = {327-343},
          doi = {10.22201/ia.01851101p.2023.59.02.12},
archivePrefix = {arXiv},
       eprint = {2308.06396},
 primaryClass = {astro-ph.GA},
       adsurl = {https://ui.adsabs.harvard.edu/abs/2023RMxAA..59..327C}
}

@ARTICLE{Storey1995,
       author = {{Storey}, P.~J. and {Hummer}, D.~G.},
        title = "{Recombination line intensities for hydrogenic ions-IV. Total recombination coefficients and machine-readable tables for Z=1 to 8}",
      journal = {\mnras},
         year = 1995,
        month = jan,
       volume = {272},
       number = {1},
        pages = {41-48},
          doi = {10.1093/mnras/272.1.41},
       adsurl = {https://ui.adsabs.harvard.edu/abs/1995MNRAS.272...41S}
}

@BOOK{Draine2011,
       author = {{Draine}, Bruce T.},
        title = "{Physics of the Interstellar and Intergalactic Medium}",
         year = 2011,
       adsurl = {https://ui.adsabs.harvard.edu/abs/2011piim.book.....D}
}

@INPROCEEDINGS{XSPEC,
       author = {{Arnaud}, K.~A.},
        title = "{XSPEC: The First Ten Years}",
    booktitle = {Astronomical Data Analysis Software and Systems V},
         year = 1996,
       editor = {{Jacoby}, George H. and {Barnes}, Jeannette},
       series = {Astronomical Society of the Pacific Conference Series},
       volume = {101},
        month = jan,
        pages = {17},
       adsurl = {https://ui.adsabs.harvard.edu/abs/1996ASPC..101...17A}
}

@BOOK{RybickiLightman1986,
       author = {{Rybicki}, George B. and {Lightman}, Alan P.},
        title = "{Radiative Processes in Astrophysics}",
         year = 1986,
       adsurl = {https://ui.adsabs.harvard.edu/abs/1986rpa..book.....R}
}

@ARTICLE{Tarter1969A,
       author = {{Tarter}, C. Bruce and {Tucker}, Wallace H. and {Salpeter}, Edwin E.},
        title = "{The Interaction of X-Ray Sources with Optically Thin Environments}",
      journal = {\apj},
         year = 1969,
        month = jun,
       volume = {156},
        pages = {943},
          doi = {10.1086/150026},
       adsurl = {https://ui.adsabs.harvard.edu/abs/1969ApJ...156..943T}
}

@ARTICLE{Tarter1969B,
       author = {{Tarter}, C. Bruce and {Salpeter}, Edwin E.},
        title = "{The Interaction of X-Ray Sources with Optically Thick Environments}",
      journal = {\apj},
         year = 1969,
        month = jun,
       volume = {156},
        pages = {953},
          doi = {10.1086/150027},
       adsurl = {https://ui.adsabs.harvard.edu/abs/1969ApJ...156..953T}
}

@ARTICLE{Metzger2014,
       author = {{Metzger}, Brian D. and {Vurm}, Indrek and {Hasco{\"e}t}, Romain and {Beloborodov}, Andrei M.},
        title = "{Ionization break-out from millisecond pulsar wind nebulae: an X-ray probe of the origin of superluminous supernovae}",
      journal = {\mnras},
         year = 2014,
        month = jan,
       volume = {437},
       number = {1},
        pages = {703-720},
          doi = {10.1093/mnras/stt1922},
archivePrefix = {arXiv},
       eprint = {1307.8115},
 primaryClass = {astro-ph.HE},
       adsurl = {https://ui.adsabs.harvard.edu/abs/2014MNRAS.437..703M}
}

@ARTICLE{Hatchett1976,
       author = {{Hatchett}, S. and {Buff}, J. and {McCray}, R.},
        title = "{Transfer of X-rays through a spherically symmetric gas cloud.}",
      journal = {\apj},
         year = 1976,
        month = jun,
       volume = {206},
        pages = {847-860},
          doi = {10.1086/154448},
       adsurl = {https://ui.adsabs.harvard.edu/abs/1976ApJ...206..847H}
}

@ARTICLE{XSTAR,
       author = {{Kallman}, T. and {Bautista}, M.},
        title = "{Photoionization and High-Density Gas}",
      journal = {\apjs},
         year = 2001,
        month = mar,
       volume = {133},
       number = {1},
        pages = {221-253},
          doi = {10.1086/319184},
       adsurl = {https://ui.adsabs.harvard.edu/abs/2001ApJS..133..221K}
}

@ARTICLE{Verner1996,
       author = {{Verner}, D.~A. and {Ferland}, G.~J. and {Korista}, K.~T. and {Yakovlev}, D.~G.},
        title = "{Atomic Data for Astrophysics. II. New Analytic Fits for Photoionization Cross Sections of Atoms and Ions}",
      journal = {\apj},
         year = 1996,
        month = jul,
       volume = {465},
        pages = {487},
          doi = {10.1086/177435},
archivePrefix = {arXiv},
       eprint = {astro-ph/9601009},
 primaryClass = {astro-ph},
       adsurl = {https://ui.adsabs.harvard.edu/abs/1996ApJ...465..487V}
}

@ARTICLE{Bell1983,
       author = {{Bell}, K.~L. and {Gilbody}, H.~B. and {Hughes}, J.~G. and {Kingston}, A.~E. and {Smith}, F.~J.},
        title = "{Recommended Data on the Electron Impact Ionization of Light Atoms and Ions}",
      journal = {Journal of Physical and Chemical Reference Data},
         year = 1983,
        month = oct,
       volume = {12},
       number = {4},
        pages = {891-916},
          doi = {10.1063/1.555700},
       adsurl = {https://ui.adsabs.harvard.edu/abs/1983JPCRD..12..891B}
}

@ARTICLE{Krolik1984,
       author = {{Krolik}, J.~H. and {Kallman}, T.~R.},
        title = "{Soft X-ray opacity in hot and photoionized gases.}",
      journal = {\apj},
         year = 1984,
        month = nov,
       volume = {286},
        pages = {366-370},
          doi = {10.1086/162608},
       adsurl = {https://ui.adsabs.harvard.edu/abs/1984ApJ...286..366K}
}

@ARTICLE{Svirski2012,
       author = {{Svirski}, Gilad and {Nakar}, Ehud and {Sari}, Re'em},
        title = "{Optical to X-Ray Supernova Light Curves Following Shock Breakout through a Thick Wind}",
      journal = {\apj},
         year = 2012,
        month = nov,
       volume = {759},
       number = {2},
          eid = {108},
        pages = {108},
          doi = {10.1088/0004-637X/759/2/108},
archivePrefix = {arXiv},
       eprint = {1202.3437},
 primaryClass = {astro-ph.HE},
       adsurl = {https://ui.adsabs.harvard.edu/abs/2012ApJ...759..108S}
}

@ARTICLE{Svirski2014,
       author = {{Svirski}, Gilad and {Nakar}, Ehud},
        title = "{Spectrum and Light Curve of a Supernova Shock Breakout through a Thick Wolf-Rayet Wind}",
      journal = {\apj},
         year = 2014,
        month = jun,
       volume = {788},
       number = {2},
          eid = {113},
        pages = {113},
          doi = {10.1088/0004-637X/788/2/113},
archivePrefix = {arXiv},
       eprint = {1402.4477},
 primaryClass = {astro-ph.HE},
       adsurl = {https://ui.adsabs.harvard.edu/abs/2014ApJ...788..113S}
}

@ARTICLE{Chevalier2012,
       author = {{Chevalier}, Roger A. and {Irwin}, Christopher M.},
        title = "{X-Rays from Supernova Shocks in Dense Mass Loss}",
      journal = {\apjl},
         year = 2012,
        month = mar,
       volume = {747},
       number = {1},
          eid = {L17},
        pages = {L17},
          doi = {10.1088/2041-8205/747/1/L17},
archivePrefix = {arXiv},
       eprint = {1201.5581},
 primaryClass = {astro-ph.HE},
       adsurl = {https://ui.adsabs.harvard.edu/abs/2012ApJ...747L..17C}
}

@ARTICLE{Wasserman2025,
       author = {{Wasserman}, Tal and {Sapir}, Nir and {Szabo}, Peter and {Waxman}, Eli},
        title = "{The Optical to X-ray Luminosity and Spectrum of Supernova Wind Breakouts}",
      journal = {arXiv e-prints},
         year = 2025,
        month = mar,
          eid = {arXiv:2504.00098},
        pages = {arXiv:2504.00098},
          doi = {10.48550/arXiv.2504.00098},
archivePrefix = {arXiv},
       eprint = {2504.00098},
 primaryClass = {astro-ph.HE},
       adsurl = {https://ui.adsabs.harvard.edu/abs/2025arXiv250400098W}
}

@ARTICLE{Margalit2022,
       author = {{Margalit}, Ben and {Quataert}, Eliot and {Ho}, Anna Y.~Q.},
        title = "{Optical to X-Ray Signatures of Dense Circumstellar Interaction in Core-collapse Supernovae}",
      journal = {\apj},
         year = 2022,
        month = apr,
       volume = {928},
       number = {2},
          eid = {122},
        pages = {122},
          doi = {10.3847/1538-4357/ac53b0},
archivePrefix = {arXiv},
       eprint = {2109.09746},
 primaryClass = {astro-ph.HE},
       adsurl = {https://ui.adsabs.harvard.edu/abs/2022ApJ...928..122M}
}

@ARTICLE{Nymark2006,
       author = {{Nymark}, T.~K. and {Fransson}, C. and {Kozma}, C.},
        title = "{X-ray emission from radiative shocks in type II supernovae}",
      journal = {\aap},
         year = 2006,
        month = apr,
       volume = {449},
       number = {1},
        pages = {171-192},
          doi = {10.1051/0004-6361:20054169},
archivePrefix = {arXiv},
       eprint = {astro-ph/0510792},
 primaryClass = {astro-ph},
       adsurl = {https://ui.adsabs.harvard.edu/abs/2006A&A...449..171N}
}

@ARTICLE{Chevalier1994,
       author = {{Chevalier}, Roger A. and {Fransson}, Claes},
        title = "{Emission from Circumstellar Interaction in Normal Type II Supernovae}",
      journal = {\apj},
         year = 1994,
        month = jan,
       volume = {420},
        pages = {268},
          doi = {10.1086/173557},
       adsurl = {https://ui.adsabs.harvard.edu/abs/1994ApJ...420..268C}
}

@ARTICLE{Ross2005,
       author = {{Ross}, R.~R. and {Fabian}, A.~C.},
        title = "{A comprehensive range of X-ray ionized-reflection models}",
      journal = {\mnras},
         year = 2005,
        month = mar,
       volume = {358},
       number = {1},
        pages = {211-216},
          doi = {10.1111/j.1365-2966.2005.08797.x},
archivePrefix = {arXiv},
       eprint = {astro-ph/0501116},
 primaryClass = {astro-ph},
       adsurl = {https://ui.adsabs.harvard.edu/abs/2005MNRAS.358..211R}
}

@ARTICLE{Levich1971,
       author = {{Levich}, E.~V. and {Syunyaev}, R.~A.},
        title = "{Heating of Gas near Quasars, Seyfert-Galaxy Nuclei, and Pulsars by Low-Frequency Radiation.}",
      journal = {\sovast},
         year = 1971,
        month = dec,
       volume = {15},
        pages = {363},
       adsurl = {https://ui.adsabs.harvard.edu/abs/1971SvA....15..363L}
}

@ARTICLE{Ercolano2003,
       author = {{Ercolano}, B. and {Barlow}, M.~J. and {Storey}, P.~J. and {Liu}, X. -W.},
        title = "{MOCASSIN: a fully three-dimensional Monte Carlo photoionization code}",
      journal = {\mnras},
         year = 2003,
        month = apr,
       volume = {340},
       number = {4},
        pages = {1136-1152},
          doi = {10.1046/j.1365-8711.2003.06371.x},
archivePrefix = {arXiv},
       eprint = {astro-ph/0209378},
 primaryClass = {astro-ph},
       adsurl = {https://ui.adsabs.harvard.edu/abs/2003MNRAS.340.1136E}
}

@ARTICLE{Dumont2000,
       author = {{Dumont}, A. -M. and {Abrassart}, A. and {Collin}, S.},
        title = "{A code for optically thick and hot photoionized media}",
      journal = {\aap},
         year = 2000,
        month = may,
       volume = {357},
        pages = {823-838},
          doi = {10.48550/arXiv.astro-ph/0003220},
archivePrefix = {arXiv},
       eprint = {astro-ph/0003220},
 primaryClass = {astro-ph},
       adsurl = {https://ui.adsabs.harvard.edu/abs/2000A&A...357..823D}
}

@ARTICLE{Svirski2014b,
       author = {{Svirski}, Gilad and {Nakar}, Ehud},
        title = "{SN 2008D: A Wolf-Rayet Explosion Through a Thick Wind}",
      journal = {\apjl},
         year = 2014,
        month = jun,
       volume = {788},
       number = {1},
          eid = {L14},
        pages = {L14},
          doi = {10.1088/2041-8205/788/1/L14},
archivePrefix = {arXiv},
       eprint = {1403.3400},
 primaryClass = {astro-ph.HE},
       adsurl = {https://ui.adsabs.harvard.edu/abs/2014ApJ...788L..14S}
}

@ARTICLE{Modjaz2009,
       author = {{Modjaz}, M. and {Li}, W. and {Butler}, N. and {Chornock}, R. and {Perley}, D. and {Blondin}, S. and {Bloom}, J.~S. and {Filippenko}, A.~V. and {Kirshner}, R.~P. and {Kocevski}, D. and {Poznanski}, D. and {Hicken}, M. and {Foley}, R.~J. and {Stringfellow}, G.~S. and {Berlind}, P. and {Barrado y Navascues}, D. and {Blake}, C.~H. and {Bouy}, H. and {Brown}, W.~R. and {Challis}, P. and {Chen}, H. and {de Vries}, W.~H. and {Dufour}, P. and {Falco}, E. and {Friedman}, A. and {Ganeshalingam}, M. and {Garnavich}, P. and {Holden}, B. and {Illingworth}, G. and {Lee}, N. and {Liebert}, J. and {Marion}, G.~H. and {Olivier}, S.~S. and {Prochaska}, J.~X. and {Silverman}, J.~M. and {Smith}, N. and {Starr}, D. and {Steele}, T.~N. and {Stockton}, A. and {Williams}, G.~G. and {Wood-Vasey}, W.~M.},
        title = "{From Shock Breakout to Peak and Beyond: Extensive Panchromatic Observations of the Type Ib Supernova 2008D Associated with Swift X-ray Transient 080109}",
      journal = {\apj},
         year = 2009,
        month = sep,
       volume = {702},
       number = {1},
        pages = {226-248},
          doi = {10.1088/0004-637X/702/1/226},
archivePrefix = {arXiv},
       eprint = {0805.2201},
 primaryClass = {astro-ph},
       adsurl = {https://ui.adsabs.harvard.edu/abs/2009ApJ...702..226M}
}

@ARTICLE{Nayana2025,
       author = {{Nayana}, A.~J. and {Margutti}, Raffaella and {Wiston}, Eli and {Chornock}, Ryan and {Campana}, Sergio and {Laskar}, Tanmoy and {Murase}, Kohta and {Krips}, Melanie and {Migliori}, Giulia and {Tsuna}, Daichi and {Alexander}, Kate D. and {Chandra}, Poonam and {Bietenholz}, Michael and {Berger}, Edo and {Chevalier}, Roger A. and {De Colle}, Fabio and {Dessart}, Luc and {Diesing}, Rebecca and {Grefenstette}, Brian W. and {Jacobson-Gal{\'a}n}, Wynn V. and {Maeda}, Keiichi and {Marcote}, Benito and {Matthews}, Daisy and {Milisavljevic}, Dan and {Ray}, Alak K. and {Reguitti}, Andrea and {Polzin}, Ava},
        title = "{Dinosaur in a Haystack: X-Ray View of the Entrails of SN 2023ixf and the Radio Afterglow of Its Interaction with the Medium Spawned by the Progenitor Star (Paper I)}",
      journal = {\apj},
         year = 2025,
        month = may,
       volume = {985},
       number = {1},
          eid = {51},
        pages = {51},
          doi = {10.3847/1538-4357/adc2fb},
archivePrefix = {arXiv},
       eprint = {2411.02647},
 primaryClass = {astro-ph.HE},
       adsurl = {https://ui.adsabs.harvard.edu/abs/2025ApJ...985...51N}
}

@ARTICLE{Soderberg2008,
       author = {{Soderberg}, A.~M. and {Berger}, E. and {Page}, K.~L. and {Schady}, P. and {Parrent}, J. and {Pooley}, D. and {Wang}, X. -Y. and {Ofek}, E.~O. and {Cucchiara}, A. and {Rau}, A. and {Waxman}, E. and {Simon}, J.~D. and {Bock}, D.~C. -J. and {Milne}, P.~A. and {Page}, M.~J. and {Barentine}, J.~C. and {Barthelmy}, S.~D. and {Beardmore}, A.~P. and {Bietenholz}, M.~F. and {Brown}, P. and {Burrows}, A. and {Burrows}, D.~N. and {Byrngelson}, G. and {Cenko}, S.~B. and {Chandra}, P. and {Cummings}, J.~R. and {Fox}, D.~B. and {Gal-Yam}, A. and {Gehrels}, N. and {Immler}, S. and {Kasliwal}, M. and {Kong}, A.~K.~H. and {Krimm}, H.~A. and {Kulkarni}, S.~R. and {Maccarone}, T.~J. and {M{\'e}sz{\'a}ros}, P. and {Nakar}, E. and {O'Brien}, P.~T. and {Overzier}, R.~A. and {de Pasquale}, M. and {Racusin}, J. and {Rea}, N. and {York}, D.~G.},
        title = "{An extremely luminous X-ray outburst at the birth of a supernova}",
      journal = {\nat},
         year = 2008,
        month = may,
       volume = {453},
       number = {7194},
        pages = {469-474},
          doi = {10.1038/nature06997},
archivePrefix = {arXiv},
       eprint = {0802.1712},
 primaryClass = {astro-ph},
       adsurl = {https://ui.adsabs.harvard.edu/abs/2008Natur.453..469S}
}

@ARTICLE{Dickinson2025,
       author = {{Dickinson}, Danielle and {Milisavljevic}, Dan and {Garretson}, Braden and {Dessart}, Luc and {Margutti}, Raffaella and {Chornock}, Ryan and {Subrayan}, Bhagya and {Hillier}, D. John and {Golub}, Eli and {Li}, Dan and {Logsdon}, Sarah E. and {Rajagopal}, Jayadev and {Ridgway}, Susan and {Smith}, Nathan and {Cynamon}, Chuck},
        title = "{The Immediate, Exemplary, and Fleeting Echelle Spectroscopy of SN 2023ixf: Monitoring Acceleration of Slow Progenitor Circumstellar Material Driven by Shock Interaction}",
      journal = {\apj},
         year = 2025,
        month = may,
       volume = {984},
       number = {1},
          eid = {71},
        pages = {71},
          doi = {10.3847/1538-4357/adc108},
archivePrefix = {arXiv},
       eprint = {2412.14406},
 primaryClass = {astro-ph.HE},
       adsurl = {https://ui.adsabs.harvard.edu/abs/2025ApJ...984...71D}
}

@ARTICLE{Zimmerman2024,
       author = {{Zimmerman}, E.~A. and {Irani}, I. and {Chen}, P. and {Gal-Yam}, A. and {Schulze}, S. and {Perley}, D.~A. and {Sollerman}, J. and {Filippenko}, A.~V. and {Shenar}, T. and {Yaron}, O. and {Shahaf}, S. and {Bruch}, R.~J. and {Ofek}, E.~O. and {De Cia}, A. and {Brink}, T.~G. and {Yang}, Y. and {Vasylyev}, S.~S. and {Ben Ami}, S. and {Aubert}, M. and {Badash}, A. and {Bloom}, J.~S. and {Brown}, P.~J. and {De}, K. and {Dimitriadis}, G. and {Fransson}, C. and {Fremling}, C. and {Hinds}, K. and {Horesh}, A. and {Johansson}, J.~P. and {Kasliwal}, M.~M. and {Kulkarni}, S.~R. and {Kushnir}, D. and {Martin}, C. and {Matuzewski}, M. and {McGurk}, R.~C. and {Miller}, A.~A. and {Morag}, J. and {Neil}, J.~D. and {Nugent}, P.~E. and {Post}, R.~S. and {Prusinski}, N.~Z. and {Qin}, Y. and {Raichoor}, A. and {Riddle}, R. and {Rowe}, M. and {Rusholme}, B. and {Sfaradi}, I. and {Sjoberg}, K.~M. and {Soumagnac}, M. and {Stein}, R.~D. and {Strotjohann}, N.~L. and {Terwel}, J.~H. and {Wasserman}, T. and {Wise}, J. and {Wold}, A. and {Yan}, L. and {Zhang}, K.},
        title = "{The complex circumstellar environment of supernova 2023ixf}",
      journal = {\nat},
         year = 2024,
        month = mar,
       volume = {627},
       number = {8005},
        pages = {759-762},
          doi = {10.1038/s41586-024-07116-6},
archivePrefix = {arXiv},
       eprint = {2310.10727},
 primaryClass = {astro-ph.HE},
       adsurl = {https://ui.adsabs.harvard.edu/abs/2024Natur.627..759Z}
}

@ARTICLE{Kasen2006,
       author = {{Kasen}, Daniel and {Thomas}, R.~C. and {Nugent}, P.},
        title = "{Time-dependent Monte Carlo Radiative Transfer Calculations for Three-dimensional Supernova Spectra, Light Curves, and Polarization}",
      journal = {\apj},
         year = 2006,
        month = nov,
       volume = {651},
       number = {1},
        pages = {366-380},
          doi = {10.1086/506190},
archivePrefix = {arXiv},
       eprint = {astro-ph/0606111},
 primaryClass = {astro-ph},
       adsurl = {https://ui.adsabs.harvard.edu/abs/2006ApJ...651..366K}
}

@ARTICLE{Roth2015,
       author = {{Roth}, Nathaniel and {Kasen}, Daniel},
        title = "{Monte Carlo Radiation-Hydrodynamics With Implicit Methods}",
      journal = {\apjs},
         year = 2015,
        month = mar,
       volume = {217},
       number = {1},
          eid = {9},
        pages = {9},
          doi = {10.1088/0067-0049/217/1/9},
archivePrefix = {arXiv},
       eprint = {1404.4652},
 primaryClass = {astro-ph.IM},
       adsurl = {https://ui.adsabs.harvard.edu/abs/2015ApJS..217....9R}
}

@ARTICLE{Netzer2008,
       author = {{Netzer}, Hagai},
        title = "{Ionized gas in active galactic nuclei}",
      journal = {\nar},
         year = 2008,
        month = aug,
       volume = {52},
       number = {6},
        pages = {257-273},
          doi = {10.1016/j.newar.2008.06.009},
       adsurl = {https://ui.adsabs.harvard.edu/abs/2008NewAR..52..257N}
}

@ARTICLE{Margutti2019,
       author = {{Margutti}, R. and {Metzger}, B.~D. and {Chornock}, R. and {Vurm}, I. and {Roth}, N. and {Grefenstette}, B.~W. and {Savchenko}, V. and {Cartier}, R. and {Steiner}, J.~F. and {Terreran}, G. and {Margalit}, B. and {Migliori}, G. and {Milisavljevic}, D. and {Alexander}, K.~D. and {Bietenholz}, M. and {Blanchard}, P.~K. and {Bozzo}, E. and {Brethauer}, D. and {Chilingarian}, I.~V. and {Coppejans}, D.~L. and {Ducci}, L. and {Ferrigno}, C. and {Fong}, W. and {G{\"o}tz}, D. and {Guidorzi}, C. and {Hajela}, A. and {Hurley}, K. and {Kuulkers}, E. and {Laurent}, P. and {Mereghetti}, S. and {Nicholl}, M. and {Patnaude}, D. and {Ubertini}, P. and {Banovetz}, J. and {Bartel}, N. and {Berger}, E. and {Coughlin}, E.~R. and {Eftekhari}, T. and {Frederiks}, D.~D. and {Kozlova}, A.~V. and {Laskar}, T. and {Svinkin}, D.~S. and {Drout}, M.~R. and {MacFadyen}, A. and {Paterson}, K.},
        title = "{An Embedded X-Ray Source Shines through the Aspherical AT 2018cow: Revealing the Inner Workings of the Most Luminous Fast-evolving Optical Transients}",
      journal = {\apj},
         year = 2019,
        month = feb,
       volume = {872},
       number = {1},
          eid = {18},
        pages = {18},
          doi = {10.3847/1538-4357/aafa01},
archivePrefix = {arXiv},
       eprint = {1810.10720},
 primaryClass = {astro-ph.HE},
       adsurl = {https://ui.adsabs.harvard.edu/abs/2019ApJ...872...18M}
}

@ARTICLE{Nayana2025b,
       author = {{Nayana}, A.~J. and {Margutti}, Raffaella and {Wiston}, Eli and {Laskar}, Tanmoy and {Migliori}, Giulia and {Chornock}, Ryan and {Galvin}, Timothy J. and {LeBaron}, Natalie and {Hajela}, Aprajita and {Christy}, Collin T. and {Sfaradi}, Itai and {Tsuna}, Daichi and {Aspegren}, Olivia and {De Colle}, Fabio and {Metzger}, Brian D. and {Lu}, Wenbin and {Beniamini}, Paz and {Kasen}, Daniel and {Berger}, Edo and {Grefenstette}, Brian W. and {Alexander}, Kate D. and {Anupama}, G.~C. and {Coppejans}, Deanne L. and {Cruz}, Luigi F. and {DeBoer}, David R and {Drout}, Maria R. and {Farah}, Wael and {Huang}, Xiaoshan and {Jacobson-Gal{\'a}n}, W.~V. and {Milisavljevic}, Dan and {Pollak}, Alexander W. and {Roth}, Nathan J. and {Sears}, Huei and {Siemion}, Andrew and {Sheikh}, Sofia Z. and {Steiner}, James F. and {Vurm}, Indrek},
        title = "{The Most Luminous Known Fast Blue Optical Transient AT 2024wpp: Unprecedented Evolution and Properties in the X-rays and Radio}",
      journal = {arXiv e-prints},
         year = 2025,
        month = aug,
          eid = {arXiv:2509.00952},
        pages = {arXiv:2509.00952},
          doi = {10.48550/arXiv.2509.00952},
archivePrefix = {arXiv},
       eprint = {2509.00952},
 primaryClass = {astro-ph.HE},
       adsurl = {https://ui.adsabs.harvard.edu/abs/2025arXiv250900952N}
}

@ARTICLE{Chevalier2008,
       author = {{Chevalier}, Roger A. and {Fransson}, Claes},
        title = "{Shock Breakout Emission from a Type Ib/c Supernova: XRT 080109/SN 2008D}",
      journal = {\apjl},
         year = 2008,
        month = aug,
       volume = {683},
       number = {2},
        pages = {L135},
          doi = {10.1086/591522},
archivePrefix = {arXiv},
       eprint = {0806.0371},
 primaryClass = {astro-ph},
       adsurl = {https://ui.adsabs.harvard.edu/abs/2008ApJ...683L.135C}
}

@ARTICLE{Mazzali2008,
       author = {{Mazzali}, Paolo A. and {Valenti}, Stefano and {Della Valle}, Massimo and {Chincarini}, Guido and {Sauer}, Daniel N. and {Benetti}, Stefano and {Pian}, Elena and {Piran}, Tsvi and {D'Elia}, Valerio and {Elias-Rosa}, Nancy and {Margutti}, Raffaella and {Pasotti}, Francesco and {Antonelli}, L. Angelo and {Bufano}, Filomena and {Campana}, Sergio and {Cappellaro}, Enrico and {Covino}, Stefano and {D'Avanzo}, Paolo and {Fiore}, Fabrizio and {Fugazza}, Dino and {Gilmozzi}, Roberto and {Hunter}, Deborah and {Maguire}, Kate and {Maiorano}, Elisabetta and {Marziani}, Paola and {Masetti}, Nicola and {Mirabel}, Felix and {Navasardyan}, Hripsime and {Nomoto}, Ken'ichi and {Palazzi}, Eliana and {Pastorello}, Andrea and {Panagia}, Nino and {Pellizza}, L.~J. and {Sari}, Re'em and {Smartt}, Stephen and {Tagliaferri}, Gianpiero and {Tanaka}, Masaomi and {Taubenberger}, Stefan and {Tominaga}, Nozomu and {Trundle}, Carrie and {Turatto}, Massimo},
        title = "{The Metamorphosis of Supernova SN 2008D/XRF 080109: A Link Between Supernovae and GRBs/Hypernovae}",
      journal = {Science},
         year = 2008,
        month = aug,
       volume = {321},
       number = {5893},
        pages = {1185},
          doi = {10.1126/science.1158088},
archivePrefix = {arXiv},
       eprint = {0807.1695},
 primaryClass = {astro-ph},
       adsurl = {https://ui.adsabs.harvard.edu/abs/2008Sci...321.1185M}
}

@ARTICLE{Dai2018,
       author = {{Dai}, Lixin and {McKinney}, Jonathan C. and {Roth}, Nathaniel and {Ramirez-Ruiz}, Enrico and {Miller}, M. Coleman},
        title = "{A Unified Model for Tidal Disruption Events}",
      journal = {\apjl},
         year = 2018,
        month = jun,
       volume = {859},
       number = {2},
          eid = {L20},
        pages = {L20},
          doi = {10.3847/2041-8213/aab429},
archivePrefix = {arXiv},
       eprint = {1803.03265},
 primaryClass = {astro-ph.HE},
       adsurl = {https://ui.adsabs.harvard.edu/abs/2018ApJ...859L..20D}
}

@ARTICLE{Auchettl2017,
       author = {{Auchettl}, Katie and {Guillochon}, James and {Ramirez-Ruiz}, Enrico},
        title = "{New Physical Insights about Tidal Disruption Events from a Comprehensive Observational Inventory at X-Ray Wavelengths}",
      journal = {\apj},
         year = 2017,
        month = apr,
       volume = {838},
       number = {2},
          eid = {149},
        pages = {149},
          doi = {10.3847/1538-4357/aa633b},
archivePrefix = {arXiv},
       eprint = {1611.02291},
 primaryClass = {astro-ph.HE},
       adsurl = {https://ui.adsabs.harvard.edu/abs/2017ApJ...838..149A}
}

@ARTICLE{Badnell2006,
       author = {{Badnell}, N.~R.},
        title = "{Radiative Recombination Data for Modeling Dynamic Finite-Density Plasmas}",
      journal = {\apjs},
         year = 2006,
        month = dec,
       volume = {167},
       number = {2},
        pages = {334-342},
          doi = {10.1086/508465},
archivePrefix = {arXiv},
       eprint = {astro-ph/0604144},
 primaryClass = {astro-ph},
       adsurl = {https://ui.adsabs.harvard.edu/abs/2006ApJS..167..334B}
}

%\begin{thebibliography}{99}
%\end{thebibliography}

%%%%%%%%%%%%%%%%%%%%%%%%%%%%%%%%%%%%%%%%%%%%%%%%%%

%%%%%%%%%%%%%%%%% APPENDICES %%%%%%%%%%%%%%%%%%%%%

\appendix
\section{Collisional ionization and Compton heating}\label{App:collisional_ion}
The analytical model discussed above assumes that the only processes that take place are photo-ionization and recombination. Our model for the Thomson thin regime is also nearly independent of the temperature of the plasma, which affects only the recombination coefficient and which we assume depends solely on $\xi$. One could expect collisional ionization to dominate over photoionization under certain conditions, or that Compton heating can significantly change the ionization state of the matter. Here, we derive these conditions and show that indeed photoionization is the dominant process, and that Compton heating becomes relevant only in cases in which the radiation will ionize its way through, regardless of the matter temperature. 
\subsection{Collisional ionization}
Collisional ionization by thermal electrons becomes efficient only when the electrons reach a temperature of a couple of times the ionization threshold energy. Assuming the matter is at this temperature, the ratio between the collisional ionization rate and the photoionization rate for $O^{7+}\to O^{8+}$ is given by: 
\begin{equation}
    \frac{n\left<\sigma_{CI}v_{th}\right>}{n_{\gamma,I} \left<\sigma_{PI}c\right>}=\frac{5\cdot10^{-11} {\rm cm^3/s}}{9.8\cdot10^{-20}\cdot c\cdot \xi}=\frac{1}{60\xi}
\end{equation}
where $\sigma_{CI}$ is the collisional ionization cross-section, given at its peak, at $\sim$2keV \citep{Bell1983} (the value of $\left<\sigma_{CI}v_{th}\right>$ is nearly constant above 2 keV), $v_{th}$ is the thermal velocity at this temperature, and we recall that the ratio of number densities of ionizing photons to electrons is $\xi$. From this relation, it is clear that we must have $\xi<\frac{1}{60}$ for collisional ionization to be dominant. However, it is difficult to get the matter to sufficiently high temperatures with such a low value of $\xi$. indeed, our simulations show that to achieve a temperature of $\sim 2 keV$, we must have at least $\xi>0.1$. This can be seen in \ref{fig:T(xi)}. 

In the reprocessing boundary case, the photon density above 10 keV can be up to $\tTin$ times higher than at 1 keV. This can heat the matter without photoionizing it. If $\xi/\tau_{in}\le \xi_c$ ($\xi$ corrected for the reprocessing boundary) is such that collisional ionization may dominate, and for the matter to be hot, we need $\xiten\gtrsim0.6$, then we need $\tTin\gtrsim 40$. We therefore limit the range of parameters in the Thomson thick case with a reprocessing boundary to this range. 
\subsection{Compton heating}
The effect of Compton heating can be read from \ref{fig:T(xi)}. This figure shows that for $\xi\lesssim 0.1-1$ (depending on $\alpha$, $\alpha=-1$ corresponds to $\xi=0.1$), the matter is always significantly below the Compton temperature, and that means that it is set by the other heating and cooling mechanisms. Recall that if $\xi \gtrsim0.1$, $W\gtrsim 0.17\cdot \left(\frac{\Sigma}{10^{24}{\rm~cm^{-2}}}\right)^{-1}\cdot \left(\frac{Z}{Z_\odot}\right)^{-1}$. This means that for $Z\gtrsim1$, Compton heating takes over when the matter is either fully ionized or requires numerical modeling. For lower values of the metallicity, a small part of the parameter space, in which the Compton temperature is relatively low, indeed receives slight Compton heating, and this is accounted for in the numerical simulations, which show our model is still approximately valid.

\section{Steady-state approximation}\label{App:steadystate}
The model above assumes a steady-state, i.e., that the system can reach ionization equilibrium on a shorter time-scale than the time over which the SED, luminosity, or column density change significantly. Here, we approximate the timescale for attaining ionization equilibrium so that this assumption can be verified.  

The relevant timescale for reaching ionization equilibrium is the recombination timescale. The $O^{8+}$ recombination time-scale is $t_{rec}\simeq\frac{1}{n \alpha_{O8\to 7}}$, where $\alpha_{O8\to 7}\simeq 1.2\cdot 10^{-11} \left(\frac{T}{10^5 {\rm  K}}\right)^{-0.7}$. Therefore, we assess the recombination time as: 
\begin{equation}
    t_{rec}\simeq 800 {\rm s}\left(\frac{n}{10^8 {\rm  
    cm^{-3}}}\right)^{-1} \left(\frac{T}{10^5 {\rm K}}\right)^{0.7}.
\end{equation}
For a wind profile, the longest timescale is set by the outer radius where the density is lowest; however, the column density and the absorption are dominated by the region between the inner radius and a few times the inner radius. 

Plugging in $T=10^6 {\rm~ K}$ and assessing the density at $\sim 10 r_\text{in}$ provides a good approximation for the timescale reported by \textsc{Cloudy} in most of the simulations. We test several simulations that report a longer timescale than the one we assess and verify that setting the age of the system to the approximation mentioned above does not change the absorption profile. 
If the system of interest has a typical timescale shorter than the recombination timescale, \textsc{Cloudy} or a similar code should be used to test the validity of approximating your system as steady-state and using the results of this work. 
In the Thomson thick cases, the diffusion time must also be short relative to the other time-scales in the system in order to attain a steady-state, and for the matter to be Compton heated, the Compton heating timescale also needs to be considered.

%%%%%%%%%%%%%%%%%%%%%%%%%%%%%%%%%%%%%%%%%%%%%%%%%%

\section{Extensions and limitations}\label{App:extensions}
The model described in \S\ref{sec: Thomson Thin} is intended for SEDs in which $L_\nu\propto\nu^{\alpha},\alpha\le 0$.  We also assume a steady-state solution. Below, we briefly describe to what extent these assumptions can be relaxed. 

\subsection{Other SEDs}
For ionizing sources with a spectrum $L_\nu \propto\nu ^{\alpha},\alpha>0$, the model above fails because most of the ionizing photons are at the high-energy cutoff rather than close to 0.5 keV. This, on one hand, leads to less efficient ionization at $\sim$ 1 keV for a given value of $\xi$, reducing the ionization at these energies; on the other hand, it could potentially lead to significant Compton heating, which can increase the collisional ionization and reduce $\alpha_B$. For these reasons, these cases must be modeled separately, defining an ionization parameter that accounts for the reduced cross-section for ionization by high-energy photons and a separate parameter for the heating. 

However, in some cases, it is reasonable to use the models discussed in this paper for rising spectra. For example, if the cutoff energy is below a few times  $ 10^7 {\rm  K}$. In this case, the ionizing radiation is localized over a small enough energy range that the spectrum is not very important. Note that if, in these cases, the metallicity is low and there is no additional UV source ionizing the helium, helium may contribute to significant absorption even at $W>1$. 

For ionizing sources with a spectrum steeper than tested, i.e., $\alpha<-2$, the analysis in this paper should generally still be valid. For these spectra, one should pay special attention to induced Compton heating, since, depending on the low-energy cutoff, it can easily lead to a high Compton temperature.  
\subsection{Induced Compton Heating}\label{sec:Induced Compton}
Including induced Compton heating, the Compton temperature is given by \citep{Levich1971}: 
\begin{equation}
    \kB T_C = \frac{h\intop\nu L_{\nu}d\nu}{4 L}+\frac{c^{3}}{32\pi}\frac{\intop\nu^{-2}L_{\nu}^{2}d\nu}{L\cdot4\pi r^{2}c}.
\end{equation}
where the first term is spontaneous Compton heating and the second is induced Compton heating, which depends strongly on the low-energy part of the spectrum. If induced Compton is negligible relative to spontaneous Compton heating, or if the Compton temperature (including induced Compton) is below $\sim 10^7~\rm K$, the analysis in the main part of this paper is valid. Otherwise, the Compton temperature considered must include the induced Compton contributions, and the portion of parameter space in which the matter is heated to the Compton temperature needs to be considered on a case-by-case basis. 
Heated matter can cause collisional ionization to dominate over photoionization and can lower the recombination coefficient, increasing the ionization fraction relative to the non-heated case.

\section{Numerical validation of the toy-model assumptions, and calibration}\label{App:Numerical validation}
\begin{figure}
    \centering
    \includegraphics[width=\columnwidth]{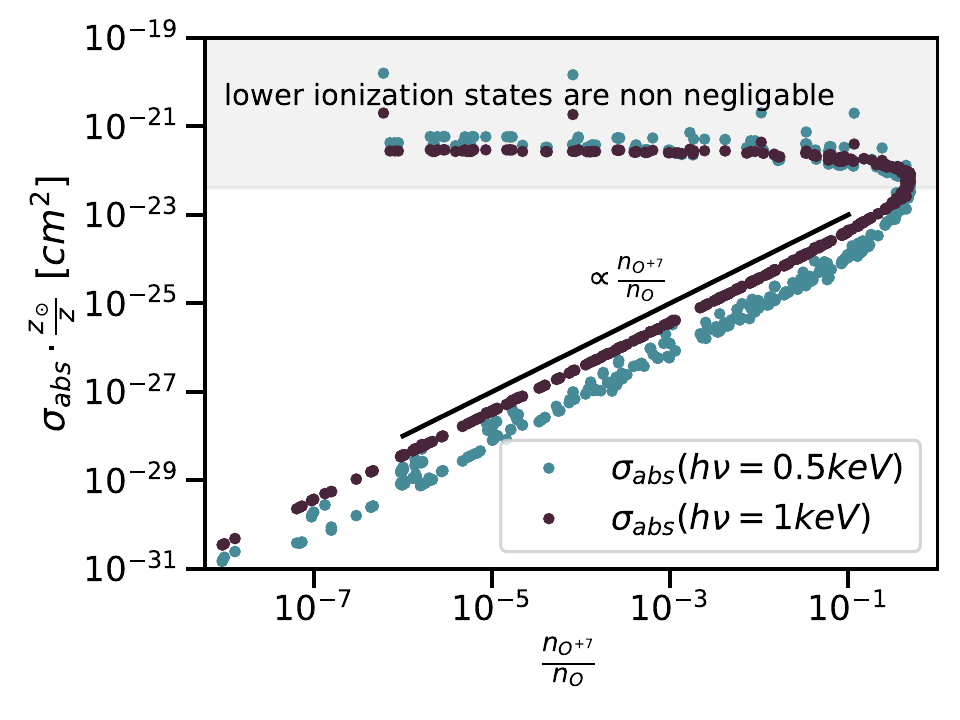}
    \caption{This figure shows the absorption cross-section at $h\nu=1~{\rm keV}$ and $h\nu=0.5~{\rm keV}$ at the illuminated side of the slab, as a function of the fraction of $O^{7+}$ - $\frac{n_{O^{7+}}}{n_
    O}$, for our complete set of simulations.}
    \label{fig:cross_section}
\end{figure}

In the main text, we make various assumptions. Here we numerically validate these assumptions. 
In figure \ref{fig:cross_section}, we validate our assumption that considering only oxygen is a good approximation for the cross-section. This figure demonstrates that for realistic abundances, the total absorption cross-section scales linearly with the oxygen ionization degree, as long as most of the oxygen is in the $O^{7+}$ and $O^{
8+}$ states. Moreover, it shows that this is a good approximation even at 0.5 keV, which is below the ionization energy of oxygen $O^{7+}$. 

\begin{figure}
    \centering
    \includegraphics[width=\columnwidth]{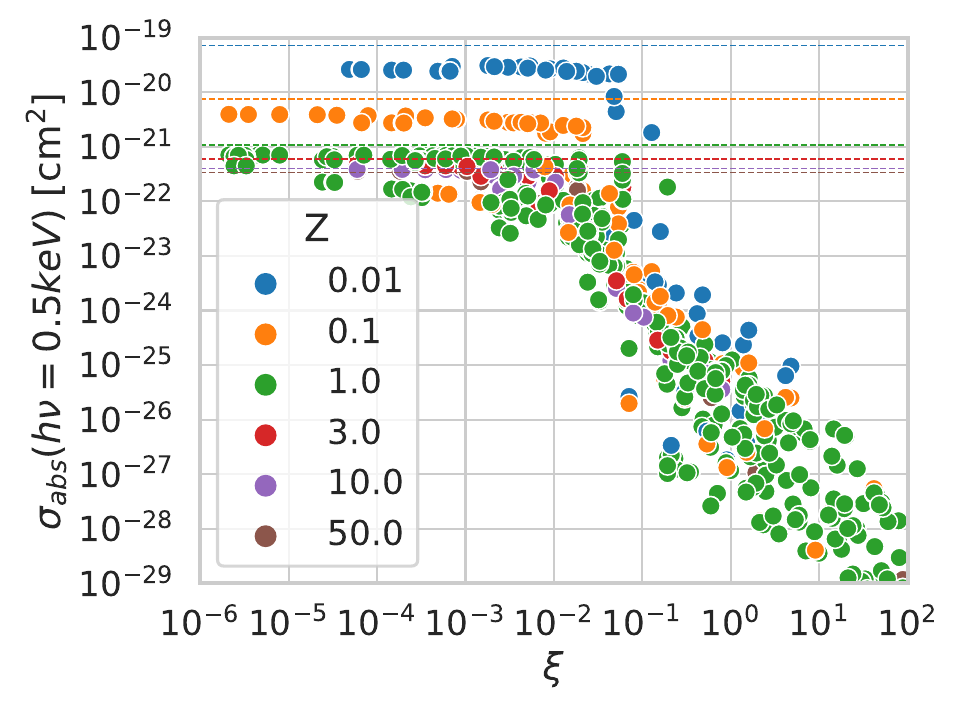}
    \caption{The absorption cross-section at 0.5 keV and the illuminated side of the cloud is plotted as a function of $\xi$. The dashed lines mark the neutral cross-section, color-coded according to metallicity.  At $\xi\lesssim\xi_c 0.015$, the cross-section varies between neutral and several times smaller than neutral, and at $\xi\gtrsim\xi_c$, the cross-section starts dropping rapidly, as the oxygen becomes fully ionized.}
    \label{fig:sigma05}
\end{figure}

Figure \ref{fig:sigma05} shows the absorption cross-section at 0.5 keV at the illuminated side of the medium scaled by metallicity, as a function of $\xi$, alongside the expected cross-section for neutral medium. Using this figure, we estimate $\xi_c\simeq 0.015$, as the critical value below which the cross-section is approximately neutral does not strongly depend on $\xi$. Note that for $\xi\ll \xi_c$, the cross-section is not perfectly neutral, since hydrogen and helium are not neutral. 

\begin{figure}
    \centering
    \includegraphics[width=\columnwidth]{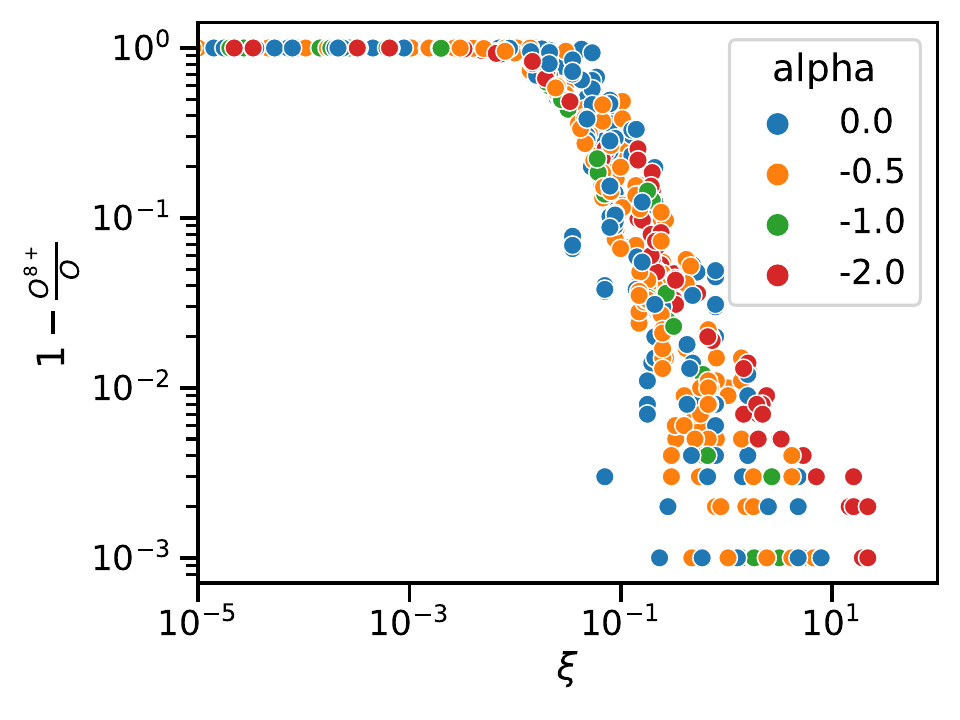}
    \caption{The non-fully ionized fraction of oxygen is plotted as a function of $\xi$. Is is clear that for $\xi\lesssim \xi_c$, none of the oxygenis in the $O^{8+}$ state, while at higher vlues of $\xi$, the $O^{8+}$ fraction grows roughly linearly. For $\xi\gtrsim0.1$, and harder spectra ($\alpha\simeq-0.5-0$), Compton heating becomes significant, reducing $\alpha_B$ in a manner that depends on the Compton temperature, and causes a significant scatter in the ionization degree.} 
    \label{fig:O8(xi)}
\end{figure}

Figure \ref{fig:O8(xi)} shows the dependance of the $O^{8+}$ fraction as a function of $\xi$. For $\xi\lesssim\xi_c$, the fraction of $O^{8+}$ is neglegible. For higher values of $\xi$, there is a roughly linear relationship between the ionization fraction and $\xi$. Once $\xi\gtrsim0.1$, the dispersion in ionization fractions grows significantly. This is primarily due to the dependance of $\alpha_B$ on the temperature. Compton heating becomes important at $\xi\sim 0.1$, leading to a change in the ionization fraction that depends on the Compton temperature, in addition to the SED slope. 

% Note, however, that if the spectrum is bluer, or has an additional high-energy component, as in the Compton thick case with a reprocessing boundary, Compton heating can dominate over line cooling when oxygen is not fully ionized, and collisional ionization can dominate over photoionization. 
\section{Monte-Carlo simulations}\label{App:Monte-Carlo}
To verify the extension to the Compton-thick medium, we perform the Monte-Carlo simulations described below. We set-up a 1D grid with a source at the origin, and an absorbing boundary at $x_{abs}$ (corresponding to the edge of the medium). Every grid cell contains $N_{\gamma}$ photons, and $N_{abs}$ absorbers. Every photon has a probability $p_{abs}$ to be absorbed by every absorber in the grid cell it crosses, and every photoionized absorber has a probability $p_{rec}$ to recombine. 

In every time step, every ionized absorber recombines with probsability $p_{rec}$, than $N_{\gamma,source}$ photons are released at the origin, then every photon takes a random step in the $+1$ or $-1$ direction, and every photon that crosses the system has a probability of $N_{abs,active}\cdot p_{abs}$ to be absorbed and ionize an absorber. Absorbed photons are not returned to the system. Then the sequence is repeated. 

We run these simulations with different boundary conditions at the origin, and track the number of absorbers, the number of photons, and the number of escaped photons in every time step, and wait until these values are converged to a steady state. We also run a similar simulation in which the photons are not scattered (always advance in the $+1$ direction). 

Comparing the reflective boundary simulations and the no-scattering simulations, we find that the same fraction of photons escapes in both cases. Comparing the simulation with an absorptive boundary to no-scattering simulations with the injected photon number decreased by $\frac{1}{x_{abs}}$, we also find a good agreement.

%%%%%%%%%%%%%%%%%%%%%%%%%%%%%%%%%%%%%%%%%%%%%%%%%%

% Don't change these lines
\bsp	% typesetting comment
\label{lastpage}
\end{document}